\documentclass[preprintnumbers, prd, twocolumn, showpacs, floatfix, preprintnumbers, 
letterpaper, 
superscriptaddress,nofootinbib]{revtex4-2}

\usepackage{amsmath}
\usepackage{amsfonts}
\usepackage{amssymb}
\usepackage{mathtools}
\usepackage{bm}
\usepackage{xcolor}
\usepackage{hyperref}
\usepackage{soul}
\usepackage[maxfloats=256]{morefloats}
\usepackage{graphicx}

\usepackage{color}
\usepackage{relsize}
\usepackage{multirow}

\newcommand{\be}{\begin{equation}}
\newcommand{\ee}{\end{equation}}
\newcommand{\bea}{\begin{eqnarray}}
\newcommand{\eea}{\end{eqnarray}}

\begin{document}

\title{Revisiting the dynamics of interacting vector-like dark energy}

\author{Carlos Rodriguez-Benites}
\email{cerodriguez@unitru.edu.pe}
\affiliation{Departamento Acad\'emico de F\'{\i}sica, Facultad de Ciencias F\'{\i}sicas y Matem\'aticas, Universidad Nacional de Trujillo, Av. Juan Pablo II s/n, Trujillo, Per\'u}
\affiliation{GRACOCC \& OASIS research groups, Facultad de Ciencias F\'{\i}sicas y Matem\'aticas, Universidad Nacional de Trujillo, Av. Juan Pablo II s/n, Trujillo, Per\'u}

\author{Manuel Gonzalez-Espinoza}
\email{manuel.gonzalez@upla.cl}
\affiliation{Laboratorio de C\'omputo de F\'isica (LCF-UPLA)\text{,} Departamento de Matem\'atica\text{,} F\'isica y Computaci\'on, Facultad de Ciencias Naturales y Exactas, Universidad de Playa Ancha, Subida Leopoldo Carvallo 270, Valpara\'iso, Chile.}
\affiliation{Laboratorio de Did\'actica de la  F\'isica, Departamento de Matem\'atica\text{,} F\'isica y Computaci\'on, Facultad de Ciencias Naturales y Exactas, Universidad de Playa Ancha, Subida Leopoldo Carvallo 270, Valpara\'iso, Chile.}

\author{Giovanni Otalora}
\email{giovanni.otalora@academicos.uta.cl}
\affiliation{Departamento de F\'isica, Facultad de Ciencias, Universidad de Tarapac\'a, Casilla 7-D, Arica, Chile}

\author{Manuel Alva-Morales}
\email{malvam@unitru.edu.pe}
\affiliation{GRACOCC \& OASIS research groups, Facultad de Ciencias F\'{\i}sicas y Matem\'aticas, Universidad Nacional de Trujillo, Av. Juan Pablo II s/n, Trujillo, Per\'u}
\affiliation{Escuela Profesional de F\'{\i}sica, Facultad de Ciencias F\'{\i}sicas y Matem\'aticas, Universidad Nacional de Trujillo, Av. Juan Pablo II s/n, Trujillo, Per\'u}

\date{\today}

\begin{abstract} 
We revise the dynamics of interacting vector-like dark energy, a theoretical framework proposed to explain the accelerated expansion of the universe. By investigating the interaction between vector-like dark energy and dark matter, we analyze its effects on the cosmic expansion history and the thermodynamics of the accelerating universe. Our results demonstrate that the presence of interaction significantly influences the evolution of vector-like dark energy, leading to distinct features in its equation of state and energy density. We compare our findings with observational data and highlight the importance of considering interactions in future cosmological studies.
\end{abstract}


\maketitle

\section{Introduction}\label{Introduction}

As revealed by observational data, the accelerated expansion of the universe is one of the most perplexing phenomena in modern cosmology. The prevailing explanation for this cosmic acceleration lies in an enigmatic entity known as dark energy, which constitutes a significant portion of the universe's energy content~\cite{ade2016planck, Copeland:2006wr}. Exploring this entity offers immense potential to unravel the mysteries surrounding the fundamental building blocks and evolution of our vast Universe. The enigma of dark energy lies in its ability to drive the universe's accelerated expansion while eluding a comprehensive understanding of its physical origin and underlying mechanisms. This cosmic force permeates the fabric of space, counteracting the gravitational pull of matter and pushing galaxies farther apart. The precise nature of dark energy remains elusive, with candidates ranging from a cosmological constant, often associated with the energy of empty space \cite{Carroll:2000fy,Padilla:2015aaa}, to dynamical scalar fields evolving over cosmic time \cite{Wetterich1995} and modified gravity theories \cite{Tsujikawa:2010zza,Nojiri:2006ri}. 

The simplest dark energy model, the $\Lambda$CDM model based on the cosmological constant and cold dark matter, faces theoretical issues like the fine-tuning problem tied to its energy scale, the cosmological constant problem, and the coincidence problem \cite{Weinberg:1988cp,Martin:2012bt}. Furthermore, notable discrepancies have emerged between the $\Lambda$CDM model predictions using cosmic microwave background (CMB) data and independent local measurements \cite{Abdalla:2022yfr,DiValentino:2020zio,DiValentino:2020vvd,Heisenberg:2022lob}. Specifically, the Hubble constant $H_{0}$ from Planck data is $4.0 \sigma$ to $6.3 \sigma$ below local measurements \cite{Riess:2019qba}, and the $\sigma_8$ clustering amplitude is higher than values from low-redshift observations \cite{Heymans:2020gsg,Nunes:2021ipq,Heisenberg:2022lob,Heisenberg:2022gqk}. Thus, these theoretical issues and recent observational findings have motivated the scientific community to explore and investigate alternative dark energy models. For instance, in the context of scalar field models and modified gravity theories, one can find in the literature non-minimally coupled scalar fields models \cite{Kasper:1988zh,Amendola:1999qq,Uzan:1999ch,Chiba:1999wt,Bartolo:1999sq,Perrotta:1999am,Gannouji:2006jm,Davari:2019tni}, coupled dark energy \cite{Ellis:1989as,Wetterich:1994bg,Damour:1993id,amendola2000coupled,DiValentino:2019ffd,DiValentino:2019jae}, $f(R,\phi)$ \cite{Tsujikawa:2007gd,DeFelice:2010gb,Faraoni:2004pi,Farajollahi:2011odw,Hammad:2017svs,SolaPeracaula:2019zsl,SolaPeracaula:2020vpg}, and $f(T,\phi)$ gravity \cite{Hohmann:2018rwf,Gonzalez-Espinoza:2020azh,Gonzalez-Espinoza:2020jss,Gonzalez-Espinoza:2021mwr,Gonzalez-Espinoza:2021qnv,Duchaniya:2022hiy,Duchaniya:2022fmc,Gonzalez-Espinoza:2023whd}, among others (see Ref. \cite{Copeland:2006wr,Nojiri:2006ri,Cai:2015emx} and references therein). 

Additionally, dark energy has not only been described using scalar fields but also through other types of fields, such as vector fields. In the context of generalized Proca theories, a massive vector field is introduced, which breaks the $U(1)$ symmetry. These vector theories belong to the class of time-like vector models. The homogeneous version of this vector field has a non-zero temporal component that is a function of cosmic time. For the most general case, its dynamics exhibit an asymptotic de Sitter attractor \cite{DeFelice:2016yws,DeFelice:2016uil,Nakamura:2019phn,DeFelice:2020icf,Cardona:2023gzq,Orjuela-Quintana:2021zoe}. However, when the field is in its canonical form and minimally coupled to gravity, the equation motion becomes trivial, lacking dynamics and rendering it incapable of acting as a source of dark energy \cite{Koivisto:2008xf}.
Furthermore, space-like vector has also been explored as a means to model inflation and dark energy \cite{Ford:1989me,Burd:1991ew,Armendariz-Picon:2004say,Koivisto:2008xf,Gomez:2020sfz,Gonzalez-Espinoza:2022hui,Rodriguez:2017wkg,Garnica:2021fuu}. While space-like models are generally associated with generating a highly anisotropic universe, there are specific scenarios where this challenge can be overcome. For instance, one of these scenarios involves the assumption of many randomly oriented vector fields, which, on average, results in an isotropic cosmological background, as described in Ref. \cite{Golovnev:2008cf,Otalora:2018bso}. Another method entails the consideration of three identical vector fields for each spatial direction, referred to as the 'cosmic triad,' a concept that also aligns with the background symmetry, as expounded upon  \cite{Armendariz-Picon:2004say}. 

Specifically, in Ref. \cite{Armendariz-Picon:2004say}, the author proposed a dark energy model based on three self-interacting vector fields minimally coupled to gravity, oriented in mutually orthogonal spatial directions, and sharing the same time-dependent length. This model, also known as the vector-like dark energy model, effectively drives the current accelerated expansion of the universe and introduces new tracking attractor solutions, making cosmic evolution insensitive to initial conditions \cite{Armendariz-Picon:2004say} (see also Refs. \cite{Koivisto:2008xf,Koivisto:2009fb,Landim:2016dxh,Gomez:2020sfz}). Furthermore, in Ref. \cite{Zhang:2009yu}, the authors examined the dark-like energy model using the Noether approach, while in Ref. \cite{Zhang:2011xea}, they investigated the unified first law within this model. 

As an interesting extension of this latter model, the Interacting Vector-like Dark Energy (IVDE) model has been proposed. For instance, in Ref. \cite{Wei:2006tn}, the authors constructed a vector-like dark energy model for which an interaction between dark energy and a background perfect fluid was assumed. They identified two types of cosmological coincidence problems: the first asks why we live in an epoch where dark energy and dust matter energy densities are comparable; the second ponders why we exist in an epoch with $w_{de}<-1$.  In this way, they found that these cosmological coincidence problems can be alleviated in such IVDE models.  Also, in Ref. \cite{Wei:2006ut}, the predictions of the IVDE model were compared with observational $H(z)$ data. In particular, they showed that the best IVDE models exhibit an oscillating feature for $H(z)$ and the equation-of-state parameter (EoS), crossing $-1$ around redshift $z\sim 1.5$.

The IVDE model offers a compelling avenue of exploration, providing a fresh approach to understanding the profound mysteries of dark energy. In this theoretical framework, dark energy is not an isolated entity but interacts with other cosmic components, such as dark matter, radiation, or neutrinos~\cite{wetterich2004phenomenological, chimento2003interacting}. These interactions, mediated by a vector-like field, can potentially influence the cosmic expansion history and the formation and evolution of large-scale structures.

Understanding the properties and dynamics of IVDE is essential for several reasons. To the present work, IVDE may offer an alternative explanation for the accelerated expansion of the universe that extends beyond the standard cosmological model. By examining specific interaction mechanisms and studying their effects on cosmic evolution, we can gain deeper insights into the nature of dark energy and its role in shaping the cosmos~\cite{boehmer2008dynamics, costa2021}. Moreover, from a broader perspective in theoretical physics, IVDE has the potential to illuminate the complex interplay between dark energy and other fundamental forces and particles. For instance, exploring IVDE within the frameworks of quantum field theory and particle physics framework could reveal significant connections, thereby bridging the gaps between our understanding of the macroscopic universe and the microscopic realm of particles and fundamental interactions \cite{Morandi:2016cet,Costa:2016tpb}.

Therefore, the investigation of IVDE represents a captivating and vital avenue to understanding the nature and properties of dark energy. Through an in-depth analysis of IVDE, we can strive towards a comprehensive understanding of the nature and properties of dark energy while also advancing our knowledge of fundamental physics and the evolution of the cosmos. This research promises to unlock new horizons in our quest to comprehend the underlying mechanisms that govern the vastness of our universe.

In this paper, we study the cosmological dynamics of IVDE, analyzing its effects on the cosmic expansion history and the thermodynamics of the accelerating universe. In particular, we have performed a detailed phase space analysis, assuming several different functional forms for the coupling function $Q$ between vector-like dark energy and dark matter.  We extend previous studies in the literature \cite{Wei:2006tn} by including not only linear functional forms but also nonlinear functions for $Q$ in terms of the energy densities. In the case of nonlinear functions, we demonstrate that the $Q$ function can exhibit sign changes throughout the cosmic evolution. 

The paper is organized as follows: In section \ref{model}, we introduce the IVDE models, present the total action of the model, and derive the field equations in a general space-time. We then obtain the cosmological equations and define the effective dark energy, along with its coupling to dark matter. In Section \ref{phspace}
we reformulate the complete set of cosmological equations using new phase-space variables to derive a closed autonomous system. Furthermore, a detailed phase-space analysis of the model is performed, where we identify the corresponding critical points and their stability conditions. In Section \ref{Num_Res}, we corroborate our analytical findings by numerically solving the field equations. In Section \ref{Thermo}, we explore the thermodynamic aspects of the interacting scenario within the context of IVDE models, determining the evolution of the temperature of dark matter and dark energy as influenced by the coupling function $Q$. Finally, in Section \ref{Remarks}, we summarize the results obtained.

\section{Vector-like Dark Energy}\label{model}
The concept of ``vector-like dark energy" refers to a ``cosmic triad," as described in reference \cite{Armendariz-Picon:2004say}. This triad consists of three identical vectors oriented in mutually orthogonal directions, ensuring the preservation of isotropy. In line with the approach presented in reference \cite{Armendariz-Picon:2004say}, we examine the scenario where vector-like dark energy is minimally coupled to gravity. The corresponding action can be expressed as follows:
\bea
&S=&\int \text{d}^{4}{x} \sqrt{-g}\Bigg[\dfrac{R}{2\kappa^2}-\sum_{a=1}^3\Big(\dfrac{1}{4}{F^a}_{\mu\nu}{F^a}^{\mu\nu}+ V({A^a}^2)\Big) \Bigg]\nonumber\\[10pt]
&&+ S_{m}+ S_{r}, \label{action00}
\eea 

where $\kappa^2=8\pi G$, with $G$ being the gravitational constant, ${F^a}_{\mu\nu}=\partial_\mu A^a_\nu-\partial_\nu A^a_\mu$ and ${A^a}^2=g^{\mu\nu}A^a_\mu A^a_\nu$. $S_m$ is the action of matter $S_r$ is the action of radiation. It is worth noting that the superscript $a$ indicates each vector field constituting the cosmic triad.

Then, varying the action \eqref{action00} with respect to the metric, we obtain the following equations:
\bea
&&\dfrac{G_{\mu\nu}}{\kappa^2}-\sum_{a=1}^3\Bigg\{{F^a}_{\mu\rho}{F^a}_\nu^\rho+2\dfrac{\text{d} V}{\text{d}{A^a}^2}A^a_\mu A^a_\nu-\nonumber\\[10pt]
&&g_{\mu\nu}\Big[\dfrac{1}{4}{F^a}_{\mu\nu}{F^a}^{\mu\nu}+ V({A^a}^2)\Big]\Bigg\}=2T^{(m)}_{\mu\nu}+2T^{(r)}_{\mu\nu}.
\eea 

And, varying with respect to the cosmic triad $A^a_\mu$ we find the equation of motion
\bea
\partial_\mu(\sqrt{-g}{F^a}^{\mu\nu})=2\sqrt{-g}\frac{\text{d}V}{\text{d}{A^a}^2}{A^a}^\nu.
\eea

Below, we detail the basic equations for the cosmic triad in a cosmological background. 

\subsection{Cosmological dynamics}\label{cosmo_dyna}

We consider a spatially flat Friedmann-Lemaître-Robertson-Walker (FLRW) universe with metric:
\bea
\text{d}{s}^2=-\text{d}{t}^2+a^2(t)\delta_{ij}\text{d}{x^i}\text{d}{x^j},
\eea

where $a$ is the scale factor, a function of the cosmic time $t$. We also assume the following ansatz:
\bea
A^a_\mu=\delta^a_\mu A(t)a(t).
\eea

Thus, the modified Friedmann equations are given by

\bea
&&\frac{3 H^2}{\kappa ^2}=\frac{3}{2}(\dot{A}+ HA)^2+3 V+\rho_m+\rho_r,\label{eq:6}\\[10pt]
&&-\frac{2 \dot{H}}{\kappa ^2}=2(\dot{A}+ HA)^2+2 A^2 V_{,A^2}+\rho_m+\frac{4}{3}\rho_r,\label{eq:7}
\eea
and the motion equation of the vector field $A$:

\bea
\Ddot{A}+3H\dot{A}+A\dot{H}+2AH^2+2AV_{,A^2}=0.
\label{eq:8}
\eea

In the above, $V\left(A^2\right)$ represents the scalar potential, $H\equiv\dot{a}/a$ stands for the Hubble rate, where a dot denotes the derivative with respect to time, and a comma indicates derivative with respect to $A^2$. Furthermore, the functions $\rho_i$ and $p_i$, with $i=m,r$ denoting either non-relativistic matter (comprising cold dark matter and baryons) or radiation, respectively, serve as the energy and pressure densities. It is important to note that in the equations mentioned earlier, we have already incorporated the respective barotropic equations of state, namely, $w_m=p_m/\rho_m=0$ and $w_r=p_r/\rho_r=1/3$.

Following Ref. \cite{Copeland:2006wr} one can rewrite the Friedmann equations \eqref{eq:6} and \eqref{eq:7} in their standard form as
\bea
&&\frac{3 H^2}{\kappa ^2}=\rho_{de}+\rho_m+\rho_r\label{frw_1},\\[10pt]
&&-\frac{2 \dot{H}}{\kappa ^2}=p_{de}+\rho_{de}+\rho_m+\frac{4}{3}\rho_r \label{frw_2}.
\eea

Therefore, we can define the effective energy and pressure densities of the model as:
\bea
&&\rho_{de}=\frac{3}{2}(\dot{A}+ HA)^2+3 V,\\[10pt]
&&p_{de}=\frac{1}{2}(\dot{A}+ HA)^2-3V+2 A^2 V_{,A^2}.
\eea

On the other hand, from the conservation of the total energy-momentum tensor we have
\bea
\dot{\rho} + 3 H(\rho + p) = 0,
\label{eq:CEM}
\eea
each component (matter, radiation, and dark energy) satisfy~\eqref{eq:CEM} separately in a scenario without interaction.And, it is worth noticing that matter and radiation have standard behavior $\rho_{m}\sim a^{-3}$ and $\rho_{r}\sim a^{-4}$, respectively.

Additionally, we can define the effective dark energy EoS as
\bea
w_{de}=\frac{p_{de}}{\rho_{de}}.
\eea

It is also convenient to introduce the total equation-of-state parameter as
\bea
w_{tot}=\frac{p_{de}+p_{r}}{\rho_{de}+\rho_m+\rho_r},
\eea

which is related to the deceleration parameter $q$ through
\bea
q=\frac{1}{2}(1+3w_{tot}),
\eea

and, hence, it is easy to see that the acceleration of the Universe occurs for $q<0$ or, equivalently, when $w_{tot}<-1/3$. 

Finally, another set of cosmological parameters that can be introduced is the standard density parameters, which are defined as
\bea
\Omega_r\equiv\frac{\kappa^2\rho_r}{3H^2}\ ,\ \Omega_m\equiv\frac{\kappa^2\rho_m}{3H^2}\ , \ \Omega_{de}\equiv\frac{\kappa^2\rho_{de}}{3H^2},
\eea

and satisfy the constraint equation
\bea
\Omega_r+\Omega_m+\Omega_{de}=1.
\eea

In what follows, we extend the previous analysis to include the interaction between vector-like dark energy and dark matter \cite{Wang:2016lxa}.

\subsection{Interacting Dark Energy}

We assume that the vector-like dark energy interacts with the dark matter through a phenomenological interaction term $Q$, according to \cite{Wang:2016lxa}
\bea
\label{rho_DE_Q}
&&\dot{\rho}_{de}+3H(\rho_{de}+p_{de})=-Q,\\[10pt]
&&\dot{\rho}_{m}+3H\rho_m=Q,
\label{rho_DM_Q} \\[10pt]
&&\dot{\rho}_{r}+4H\rho_r=0,\label{eq:16}
\eea

which preserves the total energy conservation law. By assuming this interaction between the dark components, the equation of motion \eqref{eq:8} should be changed when $Q\neq 0$ and a new term due to $Q$ will appear in its r.h.s \cite{Wei:2006tn}.

In the following section, we conduct a thorough phase-space analysis for this dark energy model. Specifically, we derive the corresponding autonomous system based on the set of cosmological equations \eqref{frw_1}, \eqref{frw_2}, \eqref{rho_DE_Q}, \eqref{rho_DM_Q} and \eqref{eq:16}.

\section{Phase-space analysis}\label{phspace}

We introduce the following set of dimensionless variables \cite{Copeland:2006wr}:
\bea
&&x=\frac{\kappa \dot{A}}{\sqrt{6}H}\ , \ y=\frac{\kappa\sqrt{V}}{\sqrt{3}H}\ , \ \lambda=-\frac{V_{,A^2}}{\kappa^2 V},\nonumber\\[10pt]
&& u=\frac{\kappa A}{\sqrt{6}}\ , \ \varrho=\frac{\kappa\sqrt{\rho_r}}{\sqrt{3}H}\ , \ \Gamma=\frac{V_{,A^2A^2}V}{(V_{,A^2})^2}, \label{dynamical_variables}
\eea

and the constraint equation:
\bea
3x^2+3y^2+3u^2+6ux+\varrho^2+\Omega_m=1.
\eea

Therefore, we obtain the dynamical system
\bea
\dfrac{\text{d}x}{\text{d}N}=&&\frac{1}{2}\Big[u^3 (3-36 \lambda  y^2)+9 u^2 x (1-4 \lambda  y^2)\nonumber\\
&&+u (9 x^2+3 (4 \lambda -3) y^2+\varrho ^2-1)\nonumber\\
&&+x(3 x^2-9 y^2+\varrho ^2-3)\Big]-\frac{Q\kappa ^2}{18 (u+x)H^3},\nonumber\\[10pt]
\dfrac{\text{d}y}{\text{d}N}=&&\frac{1}{2} y \Big[u^2 (3-36 \lambda  y^2)+6 (1-2 \lambda ) u x\nonumber\\
&&+3 x^2-9 y^2+\varrho ^2+3\Big],\nonumber\\[10pt]
\dfrac{\text{d}\varrho}{\text{d}N}=&&\frac{1}{2} \varrho  \Big[u^2 (3-36 \lambda  y^2)+6 u x+3 x^2-9y^2+\varrho ^2-1\Big],\nonumber\\[10pt]
\dfrac{\text{d}\lambda}{\text{d}N}=&&-12 (\Gamma -1) \lambda ^2 u x,\nonumber\\[10pt]
\dfrac{\text{d}u}{\text{d}N}=&&x,
\label{ODE10}
\eea

where $N\equiv\ln{a}$ is the number of e-folds, and it is used as the temporal parameter since it is a function of the cosmic scale factor. Using the above set of phase space variables, we can also write:
\bea
\Omega_{de}&=&3u^2+6ux+3x^2+3y^2,\\[10pt]
\Omega_{m}&=&1-3u^2-6ux-3x^2-3y^2-\varrho^2,\\[10pt]
\Omega_{r}&=&\varrho^2.
\eea

The equation of state (EoS) of dark energy $w_{de}=p_{de}/\rho_{de}$ can be rewritten as
\bea
w_{de} &=& \frac{u^2 (1-12 \lambda  y^2 )+2 u x+x^2-3 y^2}{3 (u^2+2 u x+x^2+y^2)},
\eea

whereas the total EoS becomes
\bea
w_{tot}&=&u^2 (1-12 \lambda  y^2)+2 u x+x^2-3 y^2+\frac{\varrho ^2}{3}.
\eea

To obtain an autonomous system from the dynamical system \eqref{ODE10}, we need to define the potentials for the vector field. From now we concentrate on the exponential potential $V(A^2)\sim e^{-\kappa^2\lambda A^2}$. Below, we study the critical points and their stability properties for various interaction cases.

In the literature, various models have been explored where the function $Q$, which represents the interaction term, is considered as a function of energy densities and the Hubble parameter \cite{Wang:2016lxa,Lepe:2017yvs}. In this paper, we use three interacting scenarios: linear, non-linear, and a sign-changeable \cite{arevalo2022dynamics}.

\subsection{Critical points}

In this section, we obtain the critical points from the conditions $\text{d}{x}/\text{d}{N}=\text{d}{y}/\text{d}{N}=\text{d}{\varrho}/\text{d}{N}=\text{d}{u}/\text{d}{N}=0$ \cite{Copeland:2006wr}. Where, if we consider the definition of each dynamical variable \eqref{dynamical_variables}, the critical points that are physically possible are given by $y_c\geq0,\ \varrho_c\geq0$ and $u_c\geq0$. \\

Consequently, we find the critical points for each interaction term: 

\subsubsection{Case I: \texorpdfstring{$Q = 3 \alpha H \rho_m$}{Q1}\label{int1}} 

Using the above set of phase space variables, we can rewrite the interaction term as:
\bea
Q= \frac{9 \alpha  H^3 (1-3 u^2-6 u x-3 x^2-3 y^2-\varrho ^2)}{\kappa ^2}.
\label{INT1}
\eea

 Critical points of the system \eqref{ODE10} with interaction case \eqref{INT1} are shown in Table \ref{table1} and the values of their cosmological parameters in Table \ref{table2}. Also, in this subsection and so on, we define $\Omega_{de}^{(r)}$ and $\Omega_{de}^{(m)}$ as the amount of dark energy during a radiation and dark matter domination era, respectively.

The critical point $a_R$ corresponds to a scaling radiation era, where $\Omega^{(r)}_{de}=u_{c}^2$. When $u_{c}=0$, we obtain the radiation-dominated solution with $\Omega_r=1$ and $w_{de}=w_{tot}=1/3$. For $u_{c}\neq 0 $, this point represents a scaling radiation era. Thus, to satisfy the early constraint imposed by the physics of big bang nucleosynthesis (BBN) and ensure $\Omega^{(r)}_{de}<0.045$ \cite{Ferreira:1997hj,Bean:2001wt}, we need to have $u_{c}<0.122$.

The critical point $c$ represents a dark energy-dominated solution with a de Sitter EoS, where $w_{de}=w_{tot}=-1$. As a result, this critical point yields accelerated expansion for all parameter values.

On the other hand, for $\alpha=0$, the critical point labeled as $b_M$ represents a matter-dominated era, with $\Omega_m = 1$ and $w_{de} = w_{tot} = 0$. In this case, the energy density of matter dominates the universe, and the EoS parameter for dark energy ($w_{de}=1/3$) and the total EoS parameter ($w_{tot}$) is zero. For $\alpha\neq 0 $, we have a scaling matter era which is constrained to satisfy  $\Omega_{de}^{(m)}< 0.02$ ($95\%$ C.L.), at redshift de $z\approx 50$, according to CMB measurements \cite{Ade:2015rim}. Thus, this leads us to the constraint  $0<\alpha<0.01$, which is also compatible with $0<\Omega_m<1$.

\begin{table}[!tbp]
 \centering
 \caption{Critical points for the autonomous system.}
\begin{center}
\begin{tabular}{c c c c c c c c c}\hline\hline
Name &  $x_c$ & $y_c$ & $\varrho_{c}$ &  $u_c$ \\ \hline
$\ \ \ \ \ \ \ \ a_{R} \ \ \ \ \ \ \ \ $ & $0$ & $0$  & $\sqrt{1-3{u_c}^2}$  & $u_c$ \\
$\ \ \ \ \ \ \ \ b_{M} \ \ \ \ \ \ \ \ $ & $0$ & $0$  & $0$  & $\sqrt{\alpha}$ \\
$\ \ \ \ \ \ \ \ c \ \ \ \ \ \ \ \ $ & $0$ & $\frac{1}{\sqrt{3\lambda}}$  & $0$  & $\sqrt{\frac{\lambda-1}{3\lambda}}$
\vspace{1.5mm}
\\ \hline\hline
\end{tabular}
\end{center}
\label{table1}
\end{table}
\begin{table}[!tbp]
 \centering
 \caption{Cosmological parameters for the critical points in Table \ref{table1}. }
\begin{center}
\begin{tabular}{c c c c c c c c c c}\hline\hline
Name &  $\Omega_{de}$ & $\Omega_m$ & $\Omega_r$ &  $w_{de}$ & $w_{tot}$ \\ \hline
$\ \ \ \ \ \ \ \ a_{R} \ \ \ \ \ \ \ \ $ & $3{u_c}^2$ & $0$  & $1-3{u_c}^2$  & $\frac{1}{3}$ & $\frac{1}{3}$\\
$\ \ \ \ \ \ \ \ b_{M} \ \ \ \ \ \ \ \ $ & $3\alpha$ & $1-3\alpha$  & $0$  & $\frac{1}{3}$ & $\alpha$ \\
$\ \ \ \ \ \ \ \ c \ \ \ \ \ \ \ \ $ & $1$ & $0$  & $0$  & $-1$ & $-1$
\vspace{1.5mm}
\\ \hline\hline
\end{tabular}
\end{center}
\label{table2}
\end{table}
\subsubsection{Case II: \texorpdfstring{$Q = 3\beta H\dfrac{{\rho_m}^2}{\rho_m+\rho_{de}}$}{Q2}\label{int2}} 
With the phase space variables mentioned earlier, we express the interaction term in the following form
\bea
Q=\frac{9 \beta  H^3 \left[3 (u+x)^2+3 y^2+\varrho ^2-1\right]^2}{\kappa ^2 \left(1-\varrho ^2\right)}.
\label{INT2}
\eea
The critical points of the system described in Equation \eqref{ODE10} under the interaction case given by Equation \eqref{INT2} are presented in Table \ref{table3}, along with the corresponding values of their cosmological parameters in Table \ref{table4}.

The critical point labeled as $d_R$ corresponds to a scaling radiation era, characterized by $\Omega^{(r)}_{de}=u_{c}^2$. When $u_{c}=0$, we obtain the radiation-dominated solution with $\Omega_r=1$ and $w_{de}=w_{tot}=1/3$. For $u_{c}\neq 0 $, this point describes a scaling radiation era.  To satisfy the early constraint imposed by the physics of big bang nucleosynthesis (BBN) and ensure $\Omega^{(r)}_{de}<0.045$ \cite{Ferreira:1997hj,Bean:2001wt}, the condition $u_{c}<0.122$ must be met.

On another note, for $\beta=0$, the critical point denoted as $e_M$ represents a matter-dominated era, where $\Omega_m = 1$ and $w_{de} = w_{tot} = 0$. In this scenario, the energy density of matter dominates the universe, resulting in a dark energy EoS parameter of $w_{de}=1/3$ and a total EoS parameter of $w_{tot}=0$. For $\beta\neq 0 $, we have a scaling matter era. Applying the constraint  $\Omega_{de}^{(m)}< 0.02$ ($95\%$ C.L.), at redshift de $z\approx 50$, according to CMB measurements \cite{Ade:2015rim}, we obtain  $0<\beta<0.01$, which is also compatible with  $0<\Omega_m<1$.

Moreover, the critical point labeled as $f$ signifies a dark energy-dominated solution characterized by a de Sitter EoS, with $w_{de}=w_{tot}=-1$. Consequently, this critical point leads to accelerated expansion regardless of the parameter values.

Finally, the points $f_1$ and $f_2$ are not physically viable, because the fractional density is constrained by $0<\Omega_m<1$, meaning that, $0<-1/\beta<1$ or $\beta <-1$, but $\beta$ must be positive and small to obtain a matter-dominated era as described by point $e_{M}$. Therefore, aiming to reproduce the standard thermal history of the universe, points $f_1$ and $f_2$ are identified as unphysical.  

\begin{table}[!tbp]
 \centering
 \caption{Critical points for the autonomous system. Where $F_{\beta}= \beta ^2 (\lambda -1)^2+\beta  \lambda  (2 \lambda +1)+\lambda ^2$ .}
\begin{center}
\begin{tabular}{c c c c c c c c c}\hline\hline
Name &  $x_c$ & $y_c$ & $\varrho_{c}$ &  $u_c$ \\ \hline
$\ d_{R} \ $ & $0$ & $0$  & $\sqrt{1-3{u_c}^2}$  & $u_c$ \\
$\ e_{M} \ $ & $0$ & $0$  & $0$  & $\sqrt{\frac{\beta}{1+3\beta}}$ \\
$\ f \ $ & $0$ & $\frac{1}{\sqrt{3\lambda}}$  & $0$  & $\sqrt{\frac{\lambda-1}{3\lambda}}$\\
$\ f_1 \ $ & $0$ & $\sqrt{\frac{\sqrt{F_{\beta}}+\beta  \lambda +\beta +\lambda }{6 \beta  \lambda }}$  & $0$  & $\sqrt{\frac{-\sqrt{F_{\beta}}+\beta  (\lambda -1)+\lambda }{6 \beta  \lambda }}$\\
$\ f_2 \ $ & $0$ & $\sqrt{\frac{-\sqrt{F_{\beta}}+\beta  \lambda +\beta +\lambda }{6 \beta  \lambda }}$  & $0$  & $\sqrt{\frac{\sqrt{F_{\beta}}+\beta  (\lambda -1)+\lambda }{6 \beta  \lambda }}$
\vspace{1.5mm}
\\ \hline\hline
\end{tabular}
\end{center}
\label{table3}
\end{table}

\begin{table}[!tbp]
 \centering
 \caption{Cosmological parameters for the critical points in Table \ref{table3}. }
\begin{center}
\begin{tabular}{c c c c c c c c c c}\hline\hline
Name &  $\Omega_{de}$ & $\Omega_m$ & $\Omega_r$ &  $w_{de}$ & $w_{tot}$ \\ \hline
$\ d_{R} \ $ & $3{u_c}^2$ & $0$  & $1-3{u_c}^2$  & $\frac{1}{3}$ & $\frac{1}{3}$\\
$\ e_{M} \ $ & $\frac{3\beta}{1+3\beta}$ & $\frac{1}{1+3\beta}$  & $0$  & $\frac{1}{3}$ & $\sqrt{\frac{\beta}{1+3\beta}}$ \\
$\ f \ $ & $1$ & $0$  & $0$  & $-1$ & $-1$\\
$\ f_1 \ $ & $\frac{1}{\beta }+1$ & $-\frac{1}{\beta }$  & $0$  & $-\frac{\beta }{\beta +1}$ & $-1$\\
$\ f_2 \ $ & $\frac{1}{\beta }+1$ & $-\frac{1}{\beta }$  & $0$  & $-\frac{\beta }{\beta +1}$ & $-1$
\vspace{1.5mm}
\\ \hline\hline
\end{tabular}
\end{center}
\label{table4}
\end{table}

\subsubsection{Case III: \texorpdfstring{$Q = \dfrac{3}{2\kappa^2}\eta H^3 (1+\Omega_r-3\Omega_{de})\Omega_m$}{Q3}\label{int3}}

Given the above set of phase space variables, we express the interaction term as follows:
\be
Q=\tfrac{3 \eta  H^3 \left(9 (u+x)^2+9 y^2-\varrho ^2-1\right) \left(3 (u+x)^2+3 y^2+\varrho ^2-1\right)}{2 \kappa ^2}.
\label{INT3}
\ee 
Critical points of the system \eqref{ODE10} with interaction case \eqref{INT3} are shown in Table \ref{table5} and the values of their cosmological parameters in Table \ref{table6}

The critical point denoted as $g_R$ corresponds to a scaling radiation era, where $\Omega^{(r)}_{de}=u_{c}^2$. When $u_{c}=0$, it gives rise to the radiation-dominated solution with $\Omega_r=1$ and $w_{de}=w_{tot}=1/3$. The case $u_{c}\neq 0 $, corresponds to a scaling radiation era.  Therefore, to meet the early constraint imposed by the physics of big bang nucleosynthesis (BBN) and ensure $\Omega^{(r)}_{de}<0.045$ \cite{Ferreira:1997hj,Bean:2001wt}, the condition $u_{c}<0.122$ must be satisfied.

Conversely, when $\eta=0$, the critical point labeled as  
$h_{M}$ represents a matter-dominated era, characterized by $\Omega_m = 1$ and $w_{de} = w_{tot} = 0$. In this scenario, the energy density of matter dominates the universe, leading to a dark energy EoS parameter of $w_{de}=1/3$. In contrast, the total equation of the state parameter remains at $w_{tot}=0$. When $\eta\neq 0$, this point represents a scaling matter era. Thus, from constraint  $\Omega_{de}^{(m)}< 0.02$ ($95\%$ C.L.), at redshift de $z\approx 50$, according to CMB measurements \cite{Ade:2015rim}, we get $-0.04<\eta<0$, which is also compatible with  $0<\Omega_m<1$. 

Additionally, the critical point $i$ corresponds to a dark energy-dominated solution with a de Sitter EoS, where $w_{de}=w_{tot}=-1$. Consequently, this critical point results in accelerated expansion for all parameter values.

Finally, points $i_{1}$ and $i_{2}$ are not physically viable. For these points, the physical condition $0<\Omega_{m}<0$ implies $\eta<-3$ or $\eta>6$, but this result is not compatible with a dark matter-dominated era as described by point $h_{M}$ ($-0.04<\eta<0$). Hence, to reproduce the standard thermal history of the universe, we designate the points $i_1$ and $i_2$ as unphysical.

\begin{table}[!tbp]
 \centering
 \caption{Critical points for the autonomous system. Where, $R_{c}=\sqrt{(2 \eta  (\lambda -3)-12 \lambda )^2-72 \eta  (\eta +3) \lambda }$ }
\begin{center}
\begin{tabular}{c c c c c c c c c}\hline\hline
Name &  $x_c$ & $y_c$ & $\varrho_{c}$ &  $u_c$ \\ \hline
$\ g_{R} \ $ & $0$ & $0$  & $\sqrt{1-3{u_c}^2}$  & $u_c$ \\
$\ h_{M} \ $ & $0$ & $0$  & $0$  & $\sqrt{\frac{\eta}{9\eta-6}}$ \\
$\ i \ $ & $0$ & $\frac{1}{\sqrt{3\lambda}}$  & $0$  & $\sqrt{\frac{\lambda-1}{3\lambda}}$\\
$\ i_1 \ $ & $0$ & $\frac{1}{6} \sqrt{\frac{2 \eta  \lambda +6 \eta -12 \lambda +R_{c}}{\eta  \lambda }}$  & $0$  & $\frac{1}{6} \sqrt{\frac{2 \eta  \lambda -6 \eta -12 \lambda -R_{c}}{\eta  \lambda }}$\\
$\ i_2 \ $ & $0$ & $\frac{1}{6} \sqrt{\frac{  2\eta \lambda +6 \eta-12 \lambda -R_{c}}{\eta  \lambda }}$  & $0$  & $\frac{1}{6} \sqrt{\frac{2 \eta  \lambda -6 \eta -12 \lambda +R_{c}}{\eta  \lambda }}$
\vspace{1.5mm}
\\ \hline\hline
\end{tabular}
\end{center}
\label{table5}
\end{table}


\begin{table}[!tbp]
 \centering
 \caption{Cosmological parameters for the critical points in Table \ref{table5}. }
\begin{center}
\begin{tabular}{c c c c c c c c c c}\hline\hline
Name &  $\Omega_{de}$ & $\Omega_m$ & $\Omega_r$ &  $w_{de}$ & $w_{tot}$ \\ \hline
$\ \ \ \ \ \ \ \ g_{R} \ \ \ \ \ \ \ \ $ & $3{u_c}^2$ & $0$  & $1-3{u_c}^2$  & $\frac{1}{3}$ & $\frac{1}{3}$\\
$\ \ \ \ \ \ \ \ h_{M} \ \ \ \ \ \ \ \ $ & $\frac{\eta }{3 \eta -2}$  & $\frac{2 (\eta -1)}{3 \eta -2}$  & $0$  & $\frac{1}{3}$ & $\frac{\eta }{9 \eta -6}$ \\
$\ \ \ \ \ \ \ \ i \ \ \ \ \ \ \ \ $ & $1$  & $0$ & $0$ & $-1$ & $-1$\\
$\ \ \ \ \ \ \ \ i_1 \ \ \ \ \ \ \ \ $ & $\frac{1}{3}-\frac{2}{\eta }$  & $\frac{2}{\eta }+\frac{2}{3}$ & $0$ & $\frac{3 \eta }{6-\eta}$ & $-1$\\
$\ \ \ \ \ \ \ \ i_2 \ \ \ \ \ \ \ \ $ & $\frac{1}{3}-\frac{2}{\eta }$  & $\frac{2}{\eta }+\frac{2}{3}$ & $0$ & $\frac{3 \eta }{6-\eta}$ & $-1$
\vspace{1.5mm}
\\ \hline\hline
\end{tabular}
\end{center}
\label{table6}
\end{table}

\begin{table*}[!t]
\caption{Properties of the critical points}
\begin{tabular}{|c|c|c|c|c|}
\hline
         Cases         & Name  & Existence & Stability  & Acceleration \\ \hline
\multirow{2}{*}{} & $a_R$ & $\forall \ \lambda , \alpha$ & unstable $\forall \ \lambda , \alpha$ & never \\ \cline{2-5} 
              I    & $b_M$ & $\alpha > 0$ & unstable for $\alpha >-1$ & never \\ \cline{2-5} 
                  & $c$ & $\lambda > 1$ & $\alpha >-1$ and $1 < \lambda \leq \frac{16}{7}$  & always \\ \hline
\multirow{2}{*}{} & $d_R$ & $\forall \ \lambda , \beta$ &  unstable $\forall \ \lambda , \beta$ & never \\ \cline{2-5}
                 & $e_M$ & $\beta<-\frac{1}{3}\lor \beta>0$ & unstable for $\beta < -\frac{1}{3}$ or $\beta>-\frac{1}{4}$ & never \\ \cline{2-5}
             II    & $f$ & $\lambda > 1$ & $1 < \lambda \leq \frac{16}{7}$ & always \\ \cline{2-5}
                 & $f_1$ & $\lambda >1\land 2 \beta +\frac{\lambda  \left(2 \lambda +\sqrt{12 \lambda -3}+1\right)}{(\lambda -1)^2}\leq 0$ & numerical solution only & always \\ \cline{2-5}
                 & $f_2$ & $(\beta >0\land \lambda \neq 0)\lor $ &  numerical solution only & always \\ 
                 & & $ \left(2 \beta +\frac{\lambda  \left(2 \lambda +\sqrt{12 \lambda -3}+1\right)}{(\lambda -1)^2}\leq 0\land \lambda >1\right)$ &  &  \\ \hline                 
\multirow{2}{*}{} & $g_R$ & $\forall \ \lambda , \eta$ & unstable $\forall \ \lambda , \eta$ & never \\ \cline{2-5}
                 & $h_M$ & $\eta <0\lor \eta >\frac{2}{3}$ & unstable for $\eta<\frac{3}{5}$ and $\eta>\frac{2}{3}$ & never \\ \cline{2-5}
                 & $i$ & $\lambda > 1$ & $\eta > -3$ and $1 < \lambda \leq \frac{16}{7}$ & always \\ \cline{2-5}
                 & & $(-3<\eta <0\land (\lambda <0\lor 0<\lambda \leq 1)) \lor$ &  &  \\
                 & & $\left(1<\lambda <3 \left(\sqrt{15}+4\right)\land \frac{3 \left(2 \lambda ^2+3 \lambda \right)}{\lambda ^2-24 \lambda +9} \right.$ & & \\ 
                 & $i_1$ & $\left. -9 \sqrt{3} \sqrt{\frac{4 \lambda ^3-\lambda ^2}{\left(\lambda ^2-24 \lambda +9\right)^2}}\leq \eta <0\right) \lor$ & numerical solution only & always\\ 
                 
             III    & & $\left(\eta <0\land \lambda =3 \left(\sqrt{15}+4\right)\right) \lor$ & & \\ 
                 & & $\left(\lambda >3 \left(\sqrt{15}+4\right)\land \right.$ & & \\ 
                 & & $\left. \left(\eta <0\lor \eta \geq \frac{3 \left(2 \lambda ^2+3 \lambda \right)}{\lambda ^2-24 \lambda +9}+9 \sqrt{3} \sqrt{\frac{4 \lambda ^3-\lambda ^2}{\left(\lambda ^2-24 \lambda +9\right)^2}}\right) \right)$ & & \\ \cline{2-5}                 
                  & & $\left(1<\lambda <3 \left(\sqrt{15}+4\right)\land \frac{3 \left(2 \lambda ^2+3 \lambda \right)}{\lambda ^2-24 \lambda +9} \right. $ &  &  \\
                  & & $\left.-9 \sqrt{3} \sqrt{\frac{4 \lambda ^3-\lambda ^2}{\left(\lambda ^2-24 \lambda +9\right)^2}}\leq \eta <-3\right)\lor $ &  &  \\                  
                  & $i_2$ & $\left(\eta <-3\land \lambda =3 \left(\sqrt{15}+4\right)\right)\lor$ & numerical solution only & always \\ 
                 & & $\left(\lambda >3 \left(\sqrt{15}+4\right)\land \right.$ & & \\
                 & & $\left. \left(\eta <-3\lor \eta \geq \frac{3 \left(2 \lambda ^2+3 \lambda \right)}{\lambda ^2-24 \lambda +9}+9 \sqrt{3} \sqrt{\frac{4 \lambda ^3-\lambda ^2}{\left(\lambda ^2-24 \lambda +9\right)^2}}\right)\right)$ & & \\  \hline
                 
\end{tabular}
\label{table7}
\end{table*}

\subsection{Stability of critical points}\label{Stability}

To study the stability of the critical points, we introduce time-dependent linear perturbations denoted as $\delta x$, $\delta y$, $\delta \varrho$, and $\delta u$ around each critical point. These perturbations take the form of $x=x_c+\delta x$, $y=y_c+\delta y$, $\varrho=\varrho_c+\delta \varrho$, and $u=u_c+\delta u$. Substituting these expressions into the autonomous system \eqref{ODE10} and linearizing the equations, we obtain the linear perturbation matrix $\mathcal{M}$, as outlined in \cite{Copeland:2006wr}. The eigenvalues of $\mathcal{M}$, denoted as $\mu_1$, $\mu_2$, $\mu_3$, and $\mu_4$, evaluated at each fixed point, determine the stability conditions for those points. Typically, the classification of stability properties proceeds as follows: (i) A \textit{stable node} exists when all the eigenvalues are negative, (ii) An \textit{unstable node} emerges when all the eigenvalues are positive, (iii) A \textit{saddle point} is characterized by having one, two, or three of the four eigenvalues as positive and the others as negative, (iv) A \textit{stable spiral} is observed when the determinant of $\mathcal{M}$ is negative, and the real part of all the eigenvalues is negative. Points classified as stable nodes or stable spirals are referred to as \textit{attractor points}, and these fixed points are reached during the cosmic evolution of the Universe, regardless of the initial conditions of the system, as long as they belong to the attraction basin of the critical point. In the following lines, we present the eigenvalues and stability conditions for each critical point in every interaction case.

\subsubsection{Case I: \texorpdfstring{$Q = 3 \alpha H \rho_m$}{Q1}}

\begin{itemize}
    \item Point $a_R$ has the eigenvalues
    \bea
    \mu_1=0\ , \ \mu_2=-1\ , \ \mu_3=2\ , \ \mu_4=1-3\alpha,
    \eea

    then this point is always unstable for all the values of $\alpha$.

    \item Point $b_M$ has the eigenvalues
    \bea
    &&\mu_1=-1\ , \ \mu_2=\frac{1}{2}(-1+3\alpha)\ , \ \mu_3=\frac{3}{2}(\alpha+1)\ , \nonumber\\[7pt]
    &&\hspace{4mm}\mu_4=-1+3\alpha,
    \eea

    which is a saddle point when $\alpha>\frac{1}{3}$ or $-1<\alpha<\frac{1}{3}$. On the other hand, it becomes a stable node when $\alpha<-1$. However, this point cannot account for the current accelerated expansion of the Universe.

    \item Point $c$ has the eigenvalues
    \bea
    &&\mu_1=-2\ , \ \mu_2=-3(\alpha+1)\, ,\nonumber\\[7pt]
    &&\hspace{4mm}\mu_{3,4}=\tfrac{-3\lambda\pm\sqrt{16\lambda-7\lambda^2}}{2\lambda}.
    \eea

    This is a de-Sitter solution, ensuring accelerated expansion for all parameter values. We observe that it is a stable node when $\alpha>-1$ and $1<\lambda\leq \frac{16}{7}$. Finally, this point never exhibits stable spiral behavior.
\end{itemize}

\begin{figure}[!tbp]
    \centering
    \includegraphics[width=0.25\textwidth]{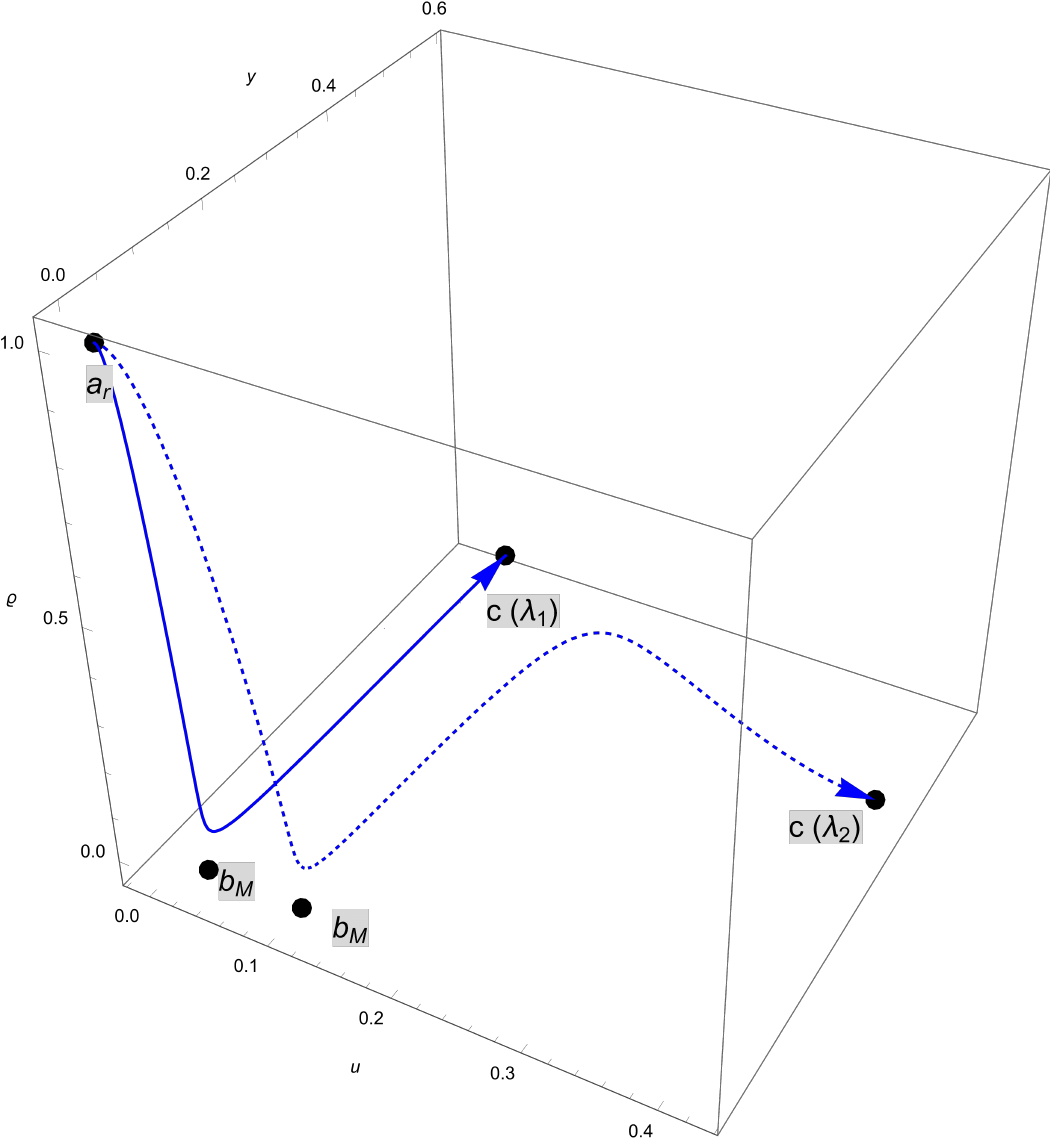}
    \caption{\scriptsize {Evolution curves in phase space for different values of $\lambda$ and $\alpha$. Blue solid lines correspond to initial conditions: $\alpha=0.0014$, $\lambda=1.004$, $x_{i}=4\times10^{-9}$, $y_{i}=4.9\times10^{-13}$, $\varrho_{i}=0.99982$, and $u_{i}=8\times10^{-5}$. Blue dashed lines correspond to initial conditions: $\alpha=0.013$, $\lambda=2.114$, $x_{i}=8\times10^{-9}$, $y_{i}=4.549\times10^{-13}$, $\varrho_{i}=0.999708$, and $u_{i}=5\times10^{-3}$. It is shown that both trajectories converge to the attractor $c$, which is a stable node that describes the dark-energy-dominated universe. Also, we have matched the current values for the fractional energy densities of dark energy $\Omega^{(0)}_{de}\approx0.68$ and dark matter $\Omega^{(0)}_{dm}\approx0.32$ , at redshift $z=0$, according to Plancks results \cite{Aghanim:2018eyx}}}
    \label{fig:A.1.7}
\end{figure}


\subsubsection{Case II: \texorpdfstring{$Q = 3\beta H\dfrac{{\rho_m}^2}{\rho_m+\rho_{de}}$}{Q2}}

\begin{itemize}
    \item Point $d_R$ has the eigenvalues
    \bea
    \mu_1=2\ , \ \mu_2=-1\ , \ \mu_3=1\ , \ \mu_4=0,
    \eea

    which tells us that it is always an unstable node.

    \item Point $e_M$ has the eigenvalues
    \bea
    &&\mu_1=-\tfrac{1}{2(1+3\beta)}\ , \ \mu_{2,3}=-1\ , \ \mu_4=\tfrac{3(1+4\beta)}{2(1+3\beta)},
    \eea

    which is a saddle point when $\beta<-\frac{1}{3}$ or $\beta>-\frac{1}{4}$. On the other hand, it becomes a stable node when $-\frac{1}{3}<\beta<-\frac{1}{4}$. However, this point cannot account for the current accelerated expansion of the Universe.

    \item Point $f$ has the eigenvalues
    \bea
    &&\mu_1=-2\ , \ \mu_2=-3\ , \ \mu_{3,4}=\tfrac{-3\pm\sqrt{-7+\frac{16}{\lambda}}}{2}.
    \eea

    This is a de-Sitter solution, hence ensuring accelerated expansion for all parameter values. We observe that it is a stable node when $1<\lambda\leq \frac{16}{7}$. Finally, this point never exhibits stable spiral behavior.

    \item Stability of points $f_1$ and $f_2$ is given by the solution of the characteristic polynomials.
    
\end{itemize}

\begin{figure}[!tbp]
    \centering
    \includegraphics[width=0.25\textwidth]{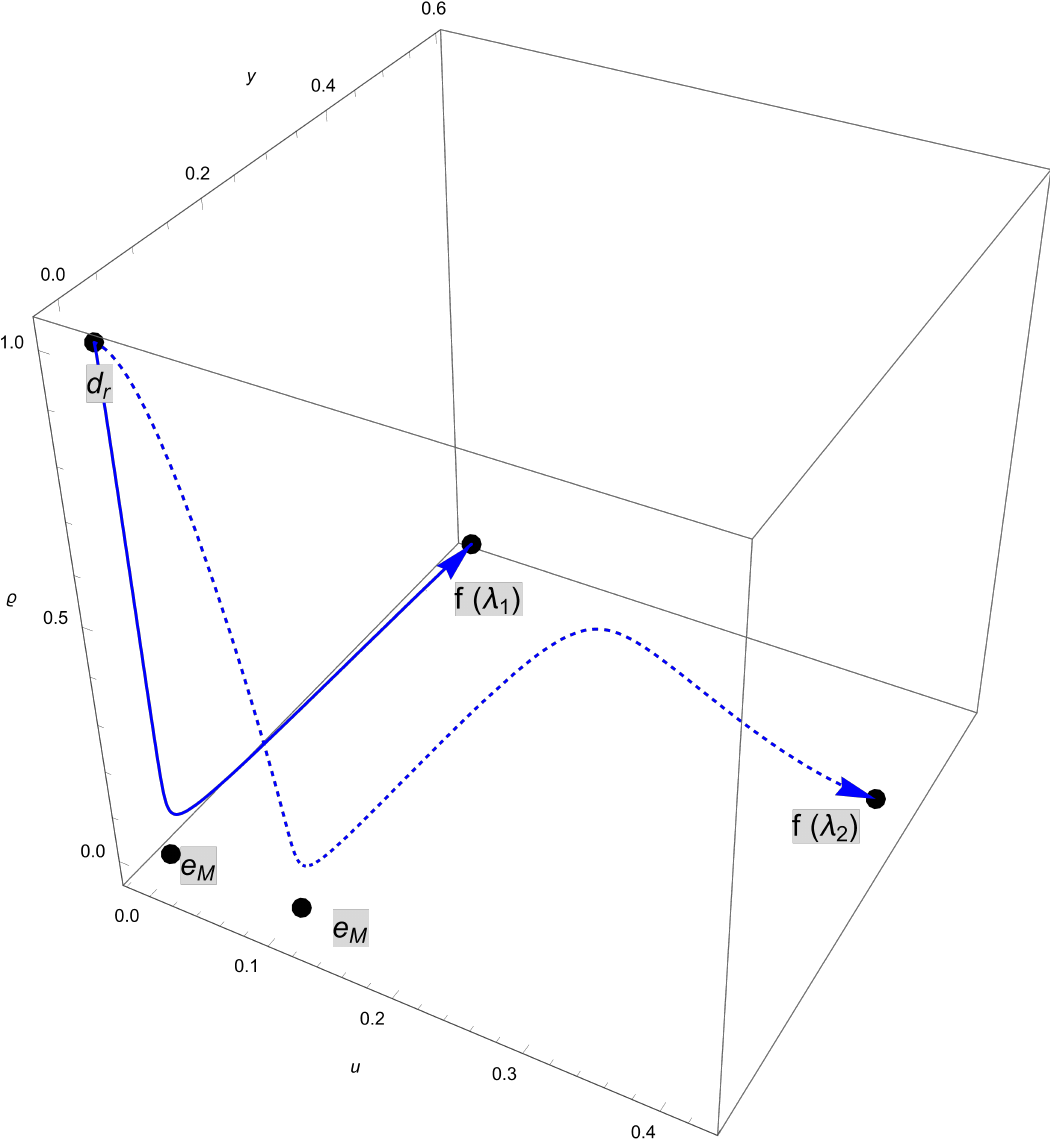}
    \caption{\scriptsize{Evolution curves in phase space for different values of $\lambda$ and $\beta$. Blue solid lines correspond to initial conditions: $\beta=0.00002$, $\lambda=1.00004$, $x_{i}=7.3\times10^{-8}$, $y_{i}=5.02\times10^{-13}$, $\varrho_{i}=0.999822$, and $u_{i}=4\times10^{-8}$. Blue dashed lines correspond to initial conditions: $\beta=0.014$, $\lambda=2.09$, $x_{i}=11.3\times10^{-7}$, $y_{i}=4.36\times10^{-13}$, $\varrho_{i}=0.999775$, and $u_{i}=10^{-5}$. It is shown that both trajectories converge to the attractor $f$, which is a stable node that describes the dark-energy-dominated universe. Also, we have matched the current values for the fractional energy densities of dark energy $\Omega^{(0)}_{de}\approx0.68$ and dark matter $\Omega^{(0)}_{dm}\approx0.32$ , at redshift $z=0$, according to Plancks results \cite{Aghanim:2018eyx}}}
    \label{fig:B.1.7}
\end{figure}



\subsubsection{Case III: \texorpdfstring{$Q = \dfrac{3}{2\kappa^2}\eta H^3 (1+\Omega_r-3\Omega_{de})\Omega_m$}{Q3}}

\begin{itemize}
    \item Point $g_R$ has the eigenvalues
    \bea
    \mu_1=0\ , \ \mu_2=-1\ , \ \mu_3=2\ , \ \mu_4=1+\eta-6\eta {u_c}^2,
    \eea

    which is always an unstable node.

    \item Point $h_M$ has the eigenvalues
    \bea
    &&\mu_1=-1\ , \ \mu_{2}=\tfrac{1-\eta}{3\eta-2}\ , \ \mu_{3}=\tfrac{3-5\eta}{2-3\eta}\ , \ \mu_4=\eta-1,
    \eea

    which acts as a saddle point when $\frac{2}{3}<\eta<1$, $\eta>1$, or $\eta<\frac{3}{5}$. Conversely, it behaves as a stable node when $\frac{3}{5}<\eta<\frac{2}{3}$. Nonetheless, this point cannot explain the current accelerated expansion of the Universe.

    \item Point $i$ has the eigenvalues
    \bea
    &&\mu_1=-2\ , \ \mu_2=-3-\eta\ , \ \mu_{3,4}=\tfrac{-3\pm\sqrt{-7+\frac{16}{\lambda}}}{2}.
    \eea

    This is a de-Sitter solution, ensuring accelerated expansion for all parameter values. We observe that it is a stable node when $\eta>-3$ and $1<\lambda\leq \frac{16}{7}$. Finally, this point never exhibits stable spiral behavior.

    \item Stability of points $i_1$ and $i_2$ is given by the solution of the characteristic polynomials.

\begin{figure}[!tbp]
    \centering
    \includegraphics[width=0.25\textwidth]{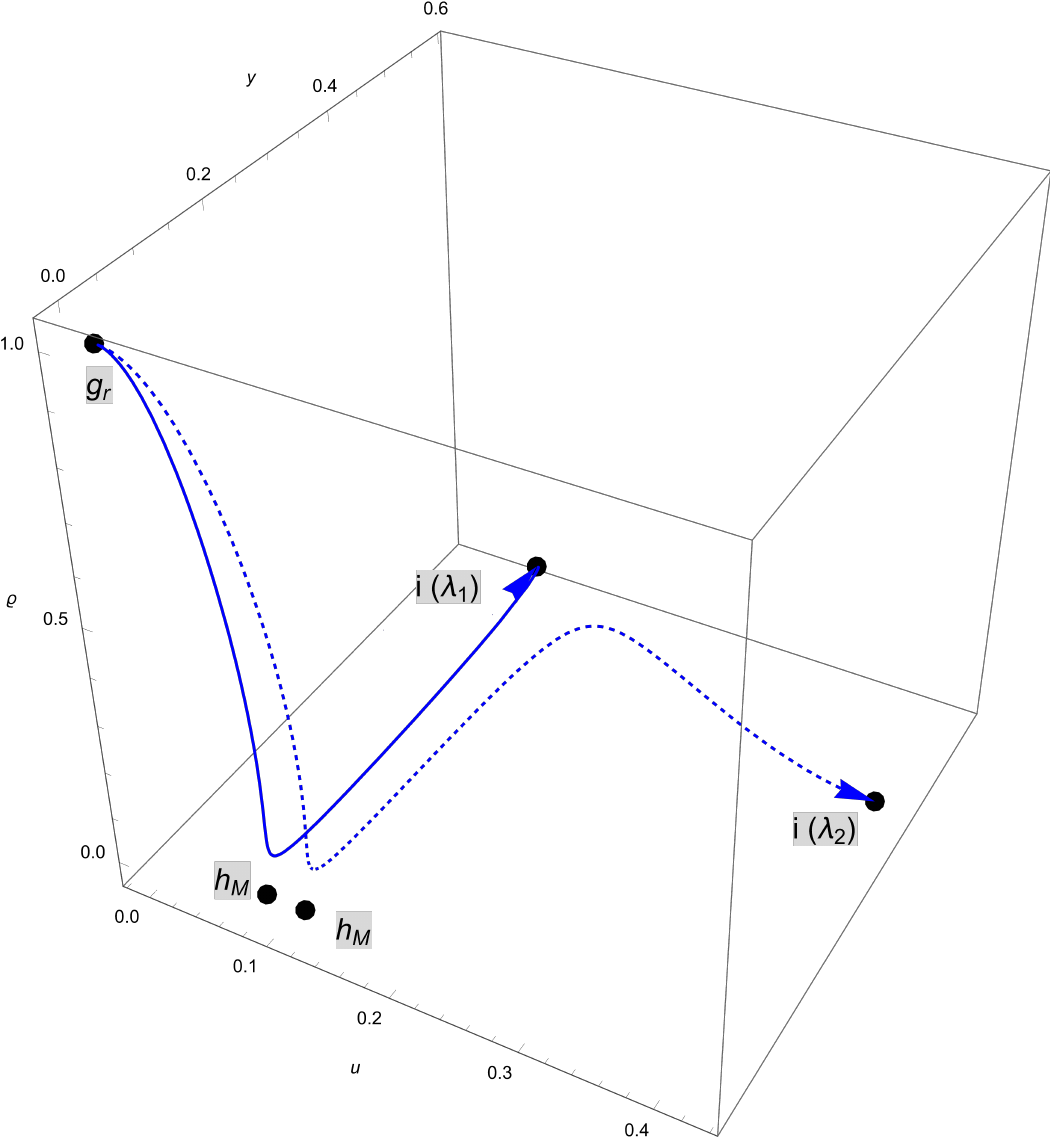}
    \caption{\scriptsize{Evolution curves in phase space for different values of $\lambda$ and $\eta$. Blue solid lines correspond to initial conditions: $\eta=-0.05$, $\lambda=1.01$, $x_{i}=10^{-7}$, $y_{i}=4.7\times10^{-13}$, $\varrho_{i}=0.99974$, and $u_{i}=10^{-7}$. Blue dashed lines correspond to initial conditions: $\eta=-0.1$, $\lambda=2.2$, $x_{i}=8\times10^{-6}$, $y_{i}=4.3\times10^{-13}$, $\varrho_{i}=0.99963$, and $u_{i}=5\times 10^{-3}$. It is shown that both trajectories converge to the attractor $i$, which is a stable node that describes the dark-energy-dominated universe. Also, we have matched the current values for the fractional energy densities of dark energy $\Omega^{(0)}_{de}\approx0.68$ and dark matter $\Omega^{(0)}_{dm}\approx0.32$ , at redshift $z=0$, according to Plancks results \cite{Aghanim:2018eyx}}}
    \label{fig:C.1.7}
\end{figure}

\end{itemize}
 
 A summary of this section is shown in Table \ref{table7}, which contains the principal properties of critical points.
 
\section{Numerical Results}\label{Num_Res}
In this section, we aim to numerically solve the autonomous system represented by equations \eqref{ODE10}, associated with a set of cosmological equations \eqref{frw_1}-\eqref{frw_2}. Our objective is to explore the characteristics of our model (for each interacting term $Q$) to explain the ongoing accelerated expansion of the Universe. Subsequently, we compare the predicted results with the most recent measurements of the Hubble parameter $H(z)$, Appendix \ref{appen_B}. We consider various parameter values and initial conditions with the aim of studying the behavior of cosmological parameters throughout the thermal history of the universe at intermediate redshifts. By investigating these several different parameter values and initial conditions, we can identify the range of potential evolutions for these cosmological parameters, all converging towards a common final-state attractor point.

\subsection{Case I: \texorpdfstring{$Q = 3 \alpha H \rho_m$}{Q1}}
\begin{figure}[!tbp]
    \centering
    \includegraphics[width=0.35\textwidth]{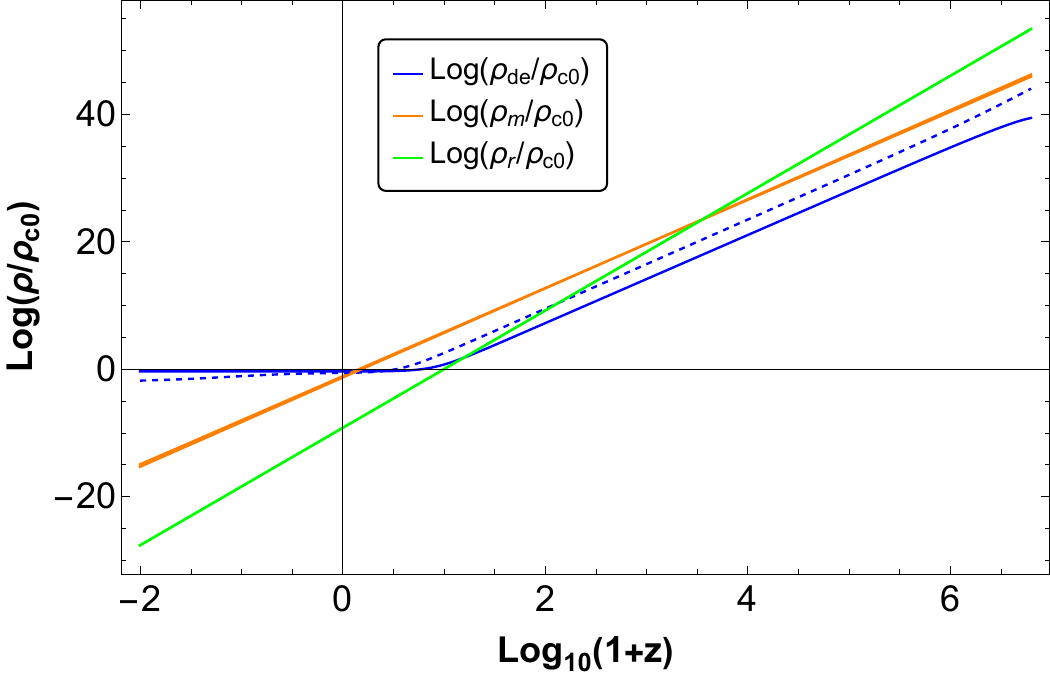}
    \caption{\scriptsize We depict the evolution of the energy density of dark energy ($\rho_{\text{de}}$) in blue, dark matter (including baryons) ($\rho_m$) in orange, and radiation ($\rho_r$) in green as functions of the redshift ($z$), for two values of $\lambda$ and $\alpha$. In particular, solid lines correspond to initial conditions: $\alpha=0.0014$, $\lambda=1.004$, $x_{i}=4\times10^{-9}$, $y_{i}=4.9\times10^{-13}$, $\varrho_{i}=0.99982$, and $u_{i}=8\times10^{-5}$. Dashed lines correspond to initial conditions: $\alpha=0.013$, $\lambda=2.114$, $x_{i}=8\times10^{-9}$, $y_{i}=4.549\times10^{-13}$, $\varrho_{i}=0.999708$, and $u_{i}=5\times10^{-3}$.} 
    \label{fig:A.1.1}
\end{figure}

\begin{figure}[!tbp]
    \centering
    \includegraphics[width=0.35\textwidth]{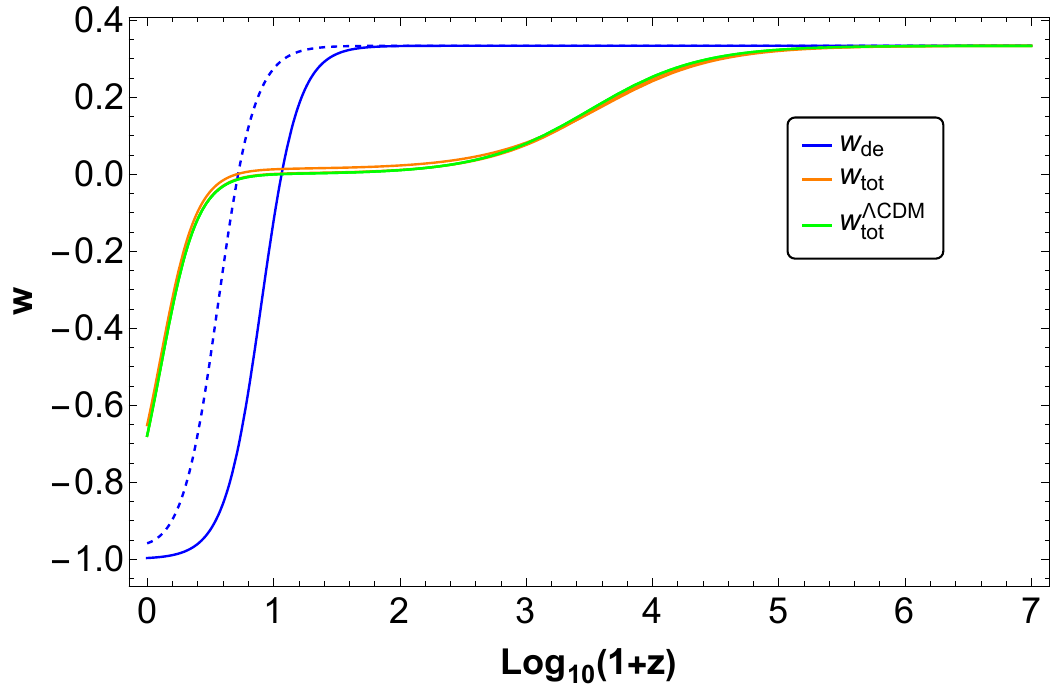}
    \caption{\scriptsize We depict the total EoS parameter $w_{\text{tot}}$ (orange line), the EOS parameter of dark energy $w_{\text{de}}$ (blue line), and the total EOS of the $\Lambda$CDM model (green line) as redshift functions. We also used the same initial conditions as shown in Figure \ref{fig:A.1.1} to obtain both the solid and dashed blue lines. For the dashed line, a value of $w_{\text{de}}\approx -0.959$ is observed at the current time $z=0$. Meanwhile, for the solid line, a value of $w_{\text{de}}\approx -0.997$ is observed at the current time $z=0$, which is consistent with the observational constraint $w_{\text{de}}^{(0)}=-1.028\pm0.032$.}.
    \label{fig:A.1.3}
\end{figure}

\begin{figure}[!tbp]
    \centering
    \includegraphics[width=0.35\textwidth]{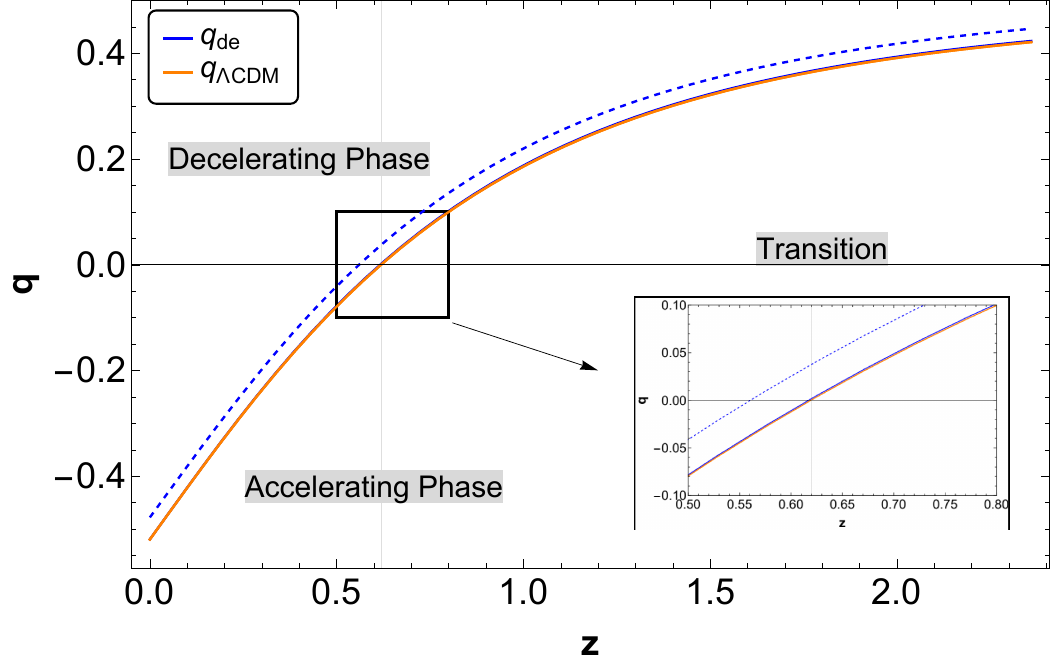}
    \caption{\scriptsize We show the evolution of the deceleration parameter $q(z)$, for the same initial conditions used in figure \ref{fig:A.1.1}.}
    \label{fig:A.1.4}
\end{figure}

\begin{figure}[!tbp]
    \centering
    \includegraphics[width=0.35\textwidth]{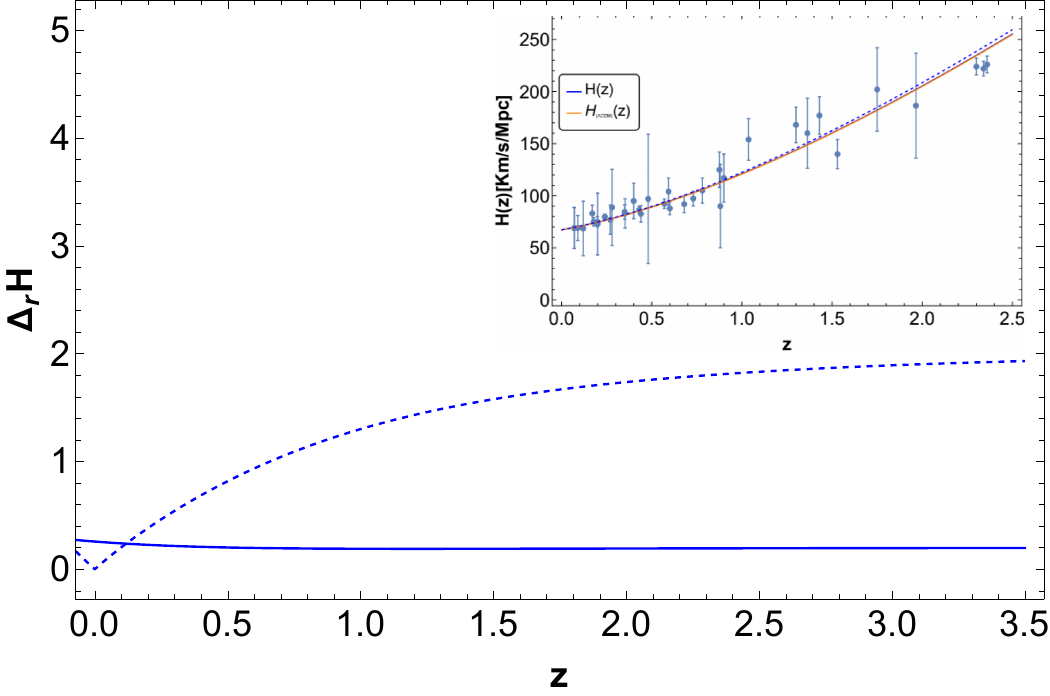}
    \caption{\scriptsize We present the evolution of the Hubble rate $H(z)$ and its relative difference $\Delta_r H(z) = 100\times |H-H_{\Lambda\text{CDM}}|/H_{\Lambda\text{CDM}}$ for the $\Lambda\text{CDM}$ model, as functions of redshift, using the same initial conditions as shown in Figure \ref{fig:A.1.1}. This is complemented by the evolution of the Hubble rate $H_{\Lambda\text{CDM}}$ in the $\Lambda$CDM model and the Hubble data from references \cite{cao2018cosmological,farooq2013hubble}. We have employed the current value of the Hubble rate, $H_0=67.4$ km/(Mpc·s), from Planck 2018 \cite{Aghanim:2018eyx}.}
    \label{fig:A.1.5}
\end{figure}

We have presented the numerical results found for this interacting model in FIGS. \ref{fig:A.1.7}, \ref{fig:A.1.1}, \ref{fig:A.1.3}, \ref{fig:A.1.4} and \ref{fig:A.1.5}.

In FIG. \ref{fig:A.1.7}, we verify that the model provides a dark energy-dominated solution with attractor behavior. The attractor is a one-parameter solution, and thus, one can have different locations for this fixed point in the field space. This implies varying temporal durations for the transitions between the decelerating and accelerating phases. However, this is restricted by cosmological observations \cite{Aghanim:2018eyx}.
In FIG \ref{fig:A.1.1}, we depict the behavior of the energy density for radiation, matter, and dark energy. Remarkably, one can observe the scaling behavior of dark energy during the radiation and matter-dominated eras. This is an essential feature of the model because scaling solutions provide a natural mechanism to alleviate the energy scale problem of dark energy \cite{Albuquerque:2018ymr,Ohashi:2009xw,Gonzalez-Espinoza:2020jss}. In FIG. \ref{fig:A.1.3}, we show the evolution of both the EoS of dark energy and the total EoS.

Additionally, we added the corresponding curve associated with the $\Lambda$CDM model to contrast the predictions of our model. During the radiation and matter-dominated eras, the EoS of dark energy behaves like a radiation field, which is an expected result since the source of dark energy is a vector field \cite{Armendariz-Picon:2004say}. Thus, in FIG. \ref{fig:A.1.4}, we observe a slight discrepancy in the deceleration-acceleration transition redshift compared to the $\Lambda$CDM result. This latter result is validated in FIG. \ref{fig:A.1.5}, where it shows the evolution of the Hubble rate $H(z)$ as a function of the redshift, along with the corresponding results from $\Lambda$CDM. In the same plot, we have depicted the relative difference for the $\Lambda$CDM model, showing that our results are compatible with observations.

\subsection{Case II: \texorpdfstring{$Q = 3\beta H\dfrac{{\rho_m}^2}{\rho_m+\rho_{de}}$}{Q2}}

\begin{figure}[!tbp]
    \centering
    \includegraphics[width=0.35\textwidth]{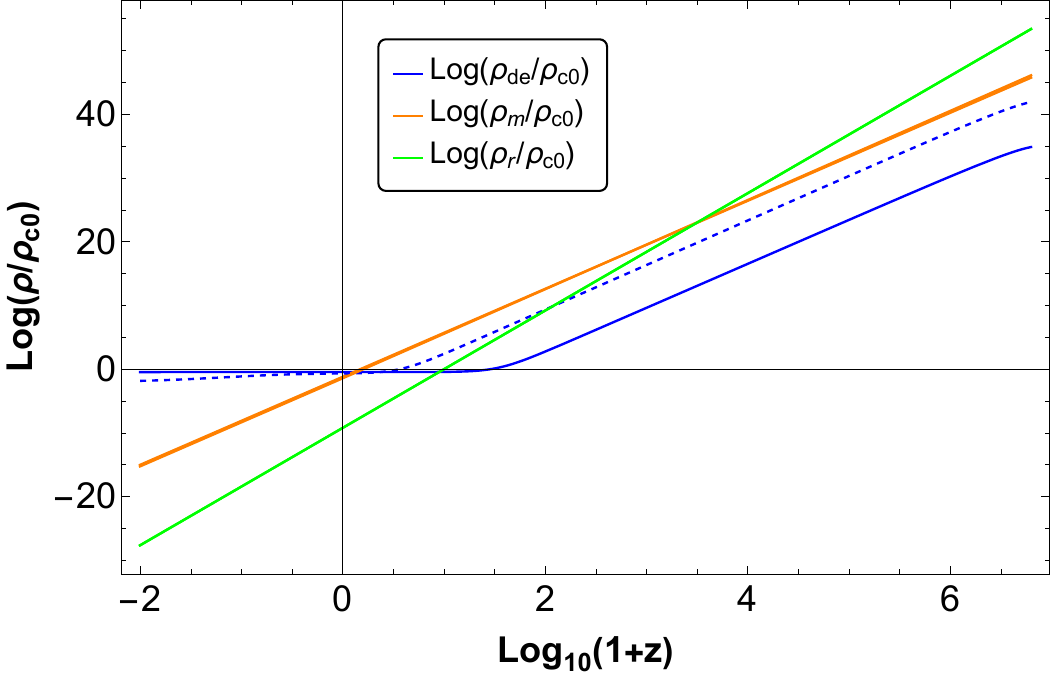}
    \caption{\scriptsize We depict the evolution of the energy density of dark energy $\rho_{\text{de}}$ (blue), dark matter (including baryons) $\rho_m$ (orange), and radiation $\rho_r$ (green) as functions of the redshift $z$, for two values of $\lambda$ and $\beta$. In particular, solid lines correspond to initial conditions: $\beta=0.00002$, $\lambda=1.00004$, $x_{i}=7.3\times10^{-8}$, $y_{i}=5.02\times10^{-13}$, $\varrho_{i}=0.999822$, and $u_{i}=4\times10^{-8}$. Dashed lines correspond to initial conditions: $\beta=0.014$, $\lambda=2.09$, $x_{i}=11.3\times10^{-7}$, $y_{i}=4.36\times10^{-13}$, $\varrho_{i}=0.999775$, and $u_{i}=10^{-5}$.} 
    \label{fig:B.1.1}
\end{figure}

\begin{figure}[!tbp]
    \centering
    \includegraphics[width=0.35\textwidth]{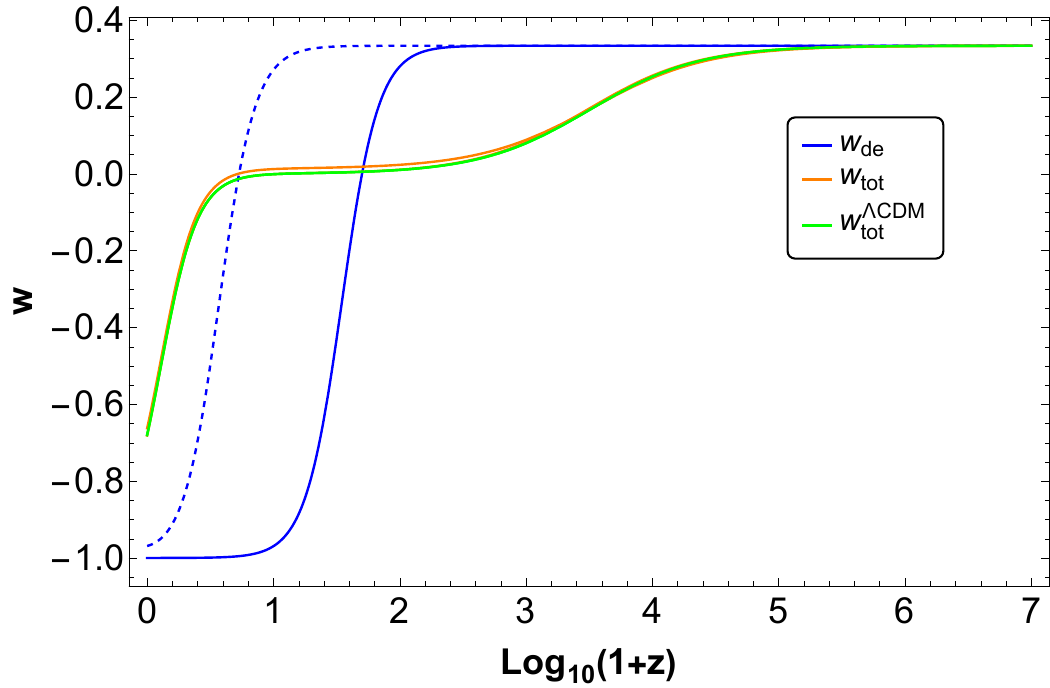}
    \caption{\scriptsize We depict the total EoS parameter $w_{\text{tot}}$ (orange line), the EOS parameter of dark energy $w_{\text{de}}$ (blue line), and the total EOS of the $\Lambda$CDM model (green line) as redshift functions. We also used the same initial conditions as shown in Figure \ref{fig:B.1.1} to obtain both the solid and dashed blue lines. For the dashed line, a value of $w_{\text{de}}\approx -0.968$ is observed at the current time $z=0$. Meanwhile, for the solid line, a value of $w_{\text{de}}\approx -0.999$ is observed at the current time $z=0$, which is consistent with the observational constraint $w_{\text{de}}^{(0)}=-1.028\pm0.032$.}
    \label{fig:B.1.3}
\end{figure}

\begin{figure}[!tbp]
    \centering
    \includegraphics[width=0.35\textwidth]{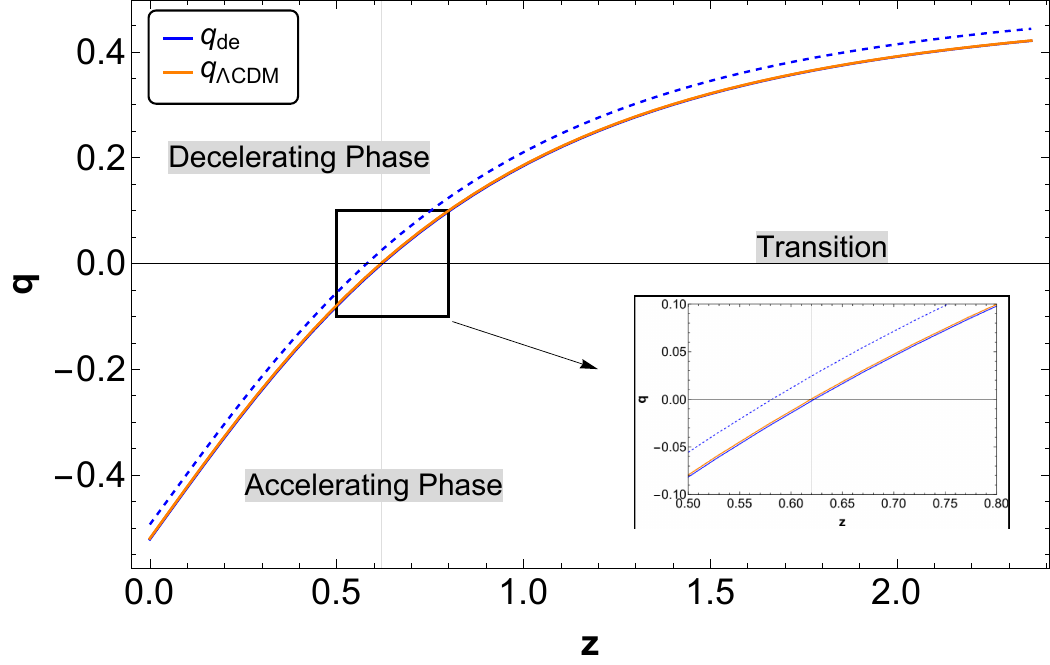}
    \caption{\scriptsize We show the evolution of the deceleration parameter $q(z)$, for the same initial conditions used in figure \ref{fig:B.1.1}.}
    \label{fig:B.1.4}
\end{figure}

\begin{figure}[!tbp]
    \centering
    \includegraphics[width=0.35\textwidth]{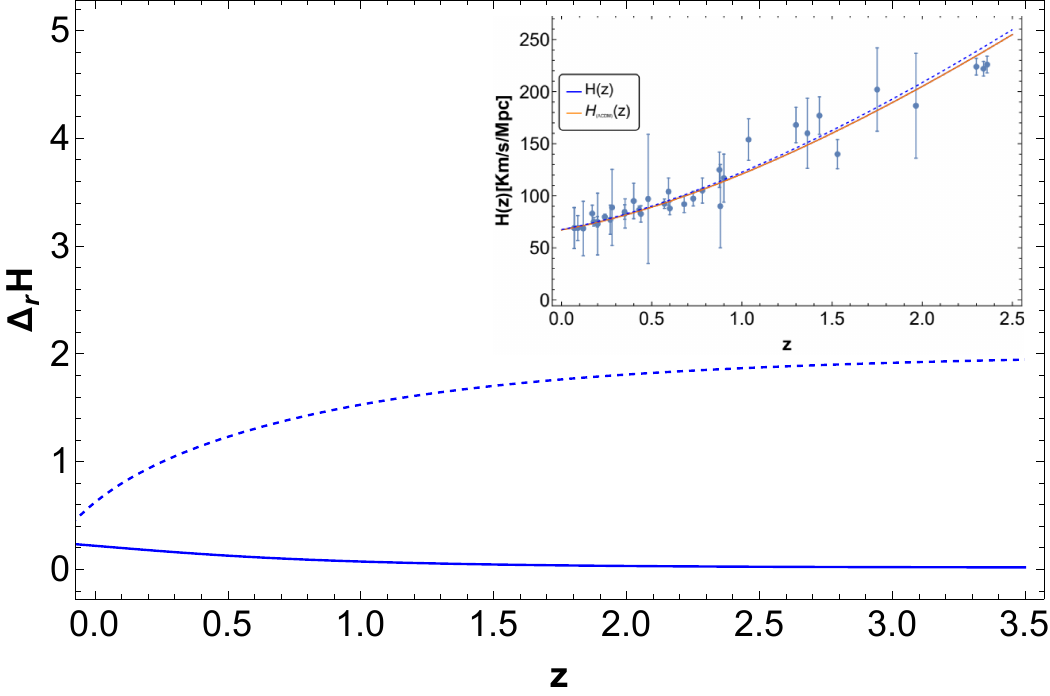}
    \caption{\scriptsize We present the evolution of the Hubble rate $H(z)$ and its relative difference $\Delta_r H(z) = 100\times |H-H_{\Lambda\text{CDM}}|/H_{\Lambda\text{CDM}}$ with respect to the $\Lambda\text{CDM}$ model, as functions of redshift, using the same initial conditions as shown in Figure \ref{fig:B.1.1}. This is complemented by the evolution of the Hubble rate $H_{\Lambda\text{CDM}}$ in the $\Lambda$CDM model and the Hubble data from references \cite{cao2018cosmological,farooq2013hubble}. We have employed the current value of the Hubble rate, $H_0=67.4$ km/(Mpc·s), from Planck 2018 \cite{Aghanim:2018eyx}.}
    \label{fig:B.1.5}
\end{figure}

For this ansatz, we have shown the numerical results in FIGS. \ref{fig:B.1.7}, \ref{fig:B.1.1}, \ref{fig:B.1.3}, \ref{fig:B.1.4} and \ref{fig:B.1.5}.

In FIG. \ref{fig:B.1.7} one can see that the system has an attractor point, which is a dark energy-dominated solution. Furthermore, this attractor point has a nature of cosmological constant with an EoS $w_{de}=-1$. However, the position of this fixed point in the phase space depends on the parameter $\lambda$. Thus, the transition time between the dark matter-dominated era and late-time acceleration depends on it, too. Let us remember that this transition time is constrained by current cosmological observations \cite{Aghanim:2018eyx}. In FIG. \ref{fig:B.1.1}, we can observe the scaling regimes during the dark-matter and radiation-dominated epochs. At early times, the dark energy component reaches higher energy scales, alleviating the energy scale problem of dark energy. In FIG. \ref{fig:B.1.3}, we depict the behavior of the EoS of dark energy and total EoS for our model, comparing the corresponding results from $\Lambda$CDM.

Interestingly enough, the evolution of the total EoS of our model is very close to that of $\Lambda$CDM, with some slight differences during the dark matter-dark energy transition time. However, the EoS of dark energy behaves like a radiation field at early times and like a cosmological constant at late times, also, in FIG. \ref{fig:B.1.4}, one can observe the decelerating-accelerating transition time as described by the evolution of the deceleration parameter. We found that this transition time is slightly displaced towards smaller redshift values than the result from $\Lambda$CDM. Nevertheless, this result is still compatible with observational data \cite{Aghanim:2018eyx}. 

Finally, in FIG. \ref{fig:B.1.5}, we numerically corroborated that our model is compatible with observational data.  The evolution of the Hubble rate is consistent with Hubble rate data found in the literature \cite{cao2018cosmological,farooq2013hubble}.

\subsection{Case III: \texorpdfstring{$Q = \dfrac{3}{2\kappa^2}\eta H^3 (1+\Omega_r-3\Omega_{de})\Omega_m$}{Q3}}

\begin{figure}[!tbp]
    \centering
    \includegraphics[width=0.35\textwidth]{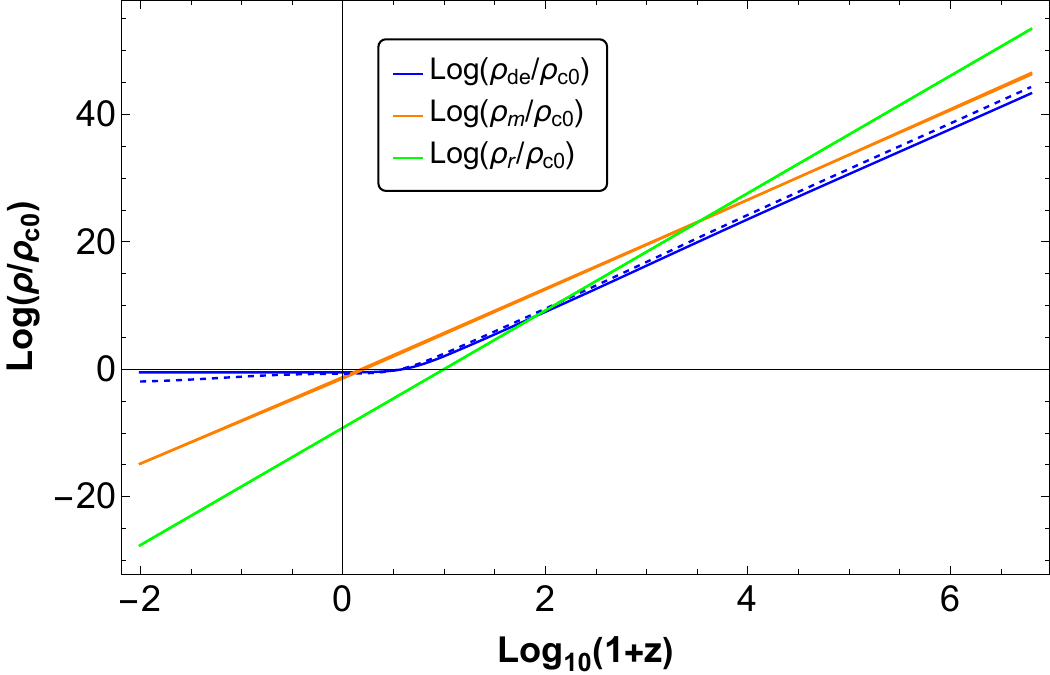}
    \caption{\scriptsize We depict the evolution of the energy density of dark energy $\rho_{\text{de}}$ (blue), dark matter (including baryons) $\rho_m$ (orange), and radiation $\rho_r$ (green) as functions of the redshift $z$, for two values of $\lambda$ and $\beta$. In particular, solid lines correspond to initial conditions: $\eta=-0.05$, $\lambda=1.01$, $x_{i}=10^{-7}$, $y_{i}=4.7\times10^{-13}$, $\varrho_{i}=0.99974$, and $u_{i}=10^{-7}$. Dashed lines correspond to initial conditions: $\eta=-0.1$, $\lambda=2.2$, $x_{i}=8\times10^{-6}$, $y_{i}=4.3\times10^{-13}$, $\varrho_{i}=0.99963$, and $u_{i}=5\times 10^{-3}$.} 
    \label{fig:C.1.1}
\end{figure}

\begin{figure}[!tbp]
    \centering
    \includegraphics[width=0.35\textwidth]{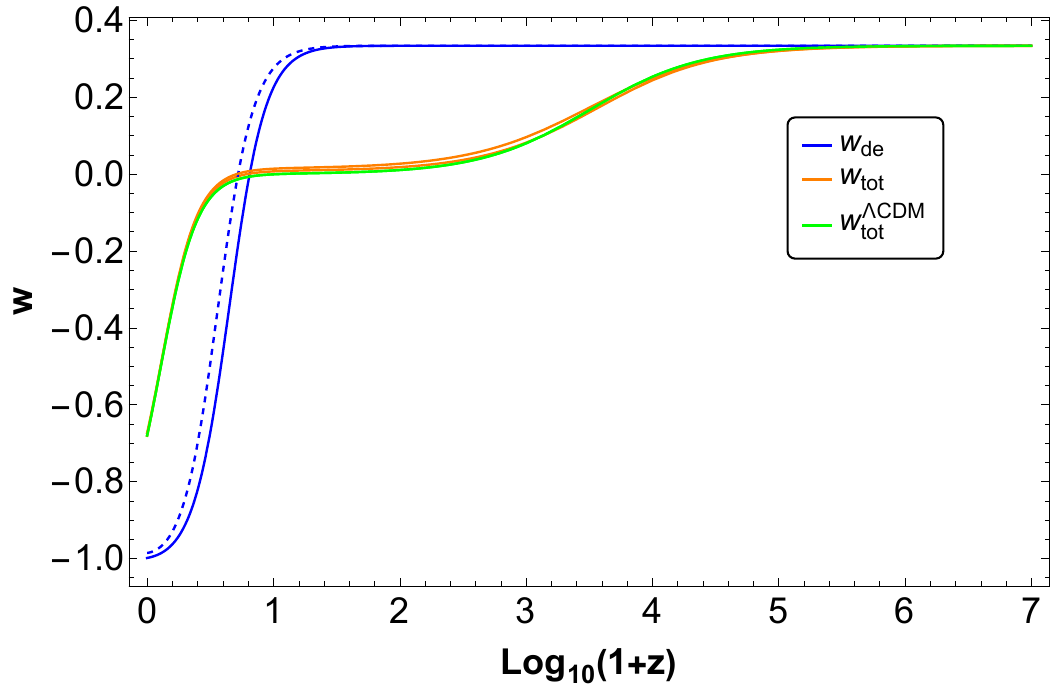}
    \caption{\scriptsize We depict the total EoS parameter $w_{\text{tot}}$ (orange line), the EOS parameter of dark energy $w_{\text{de}}$ (blue line), and the total EOS of the $\Lambda$CDM model (green line) as redshift functions. We also used the same initial conditions as shown in Figure \ref{fig:C.1.1} to obtain both the solid and dashed blue lines. For the dashed line, a value of $w_{\text{de}}\approx -0.986$ is observed at the current time $z=0$. Meanwhile, for the solid line, a value of $w_{\text{de}}\approx -0.999$ is observed at the current time $z=0$, which is consistent with the observational constraint $w_{\text{de}}^{(0)}=-1.028\pm0.032$.}
    \label{fig:C.1.3}
\end{figure}

\begin{figure}[!tbp]
    \centering
    \includegraphics[width=0.35\textwidth]{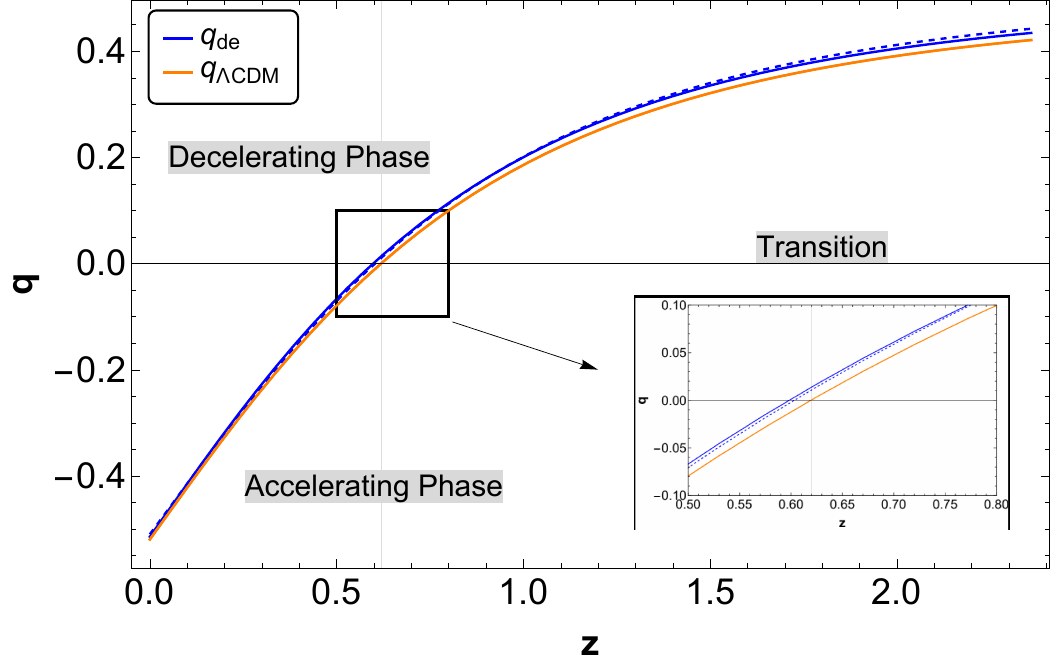}
    \caption{\scriptsize We show the evolution of the deceleration parameter $q(z)$, for the same initial conditions used in figure \ref{fig:C.1.1}.}
    \label{fig:C.1.4}
\end{figure}

\begin{figure}[!tbp]
    \centering
    \includegraphics[width=0.35\textwidth]{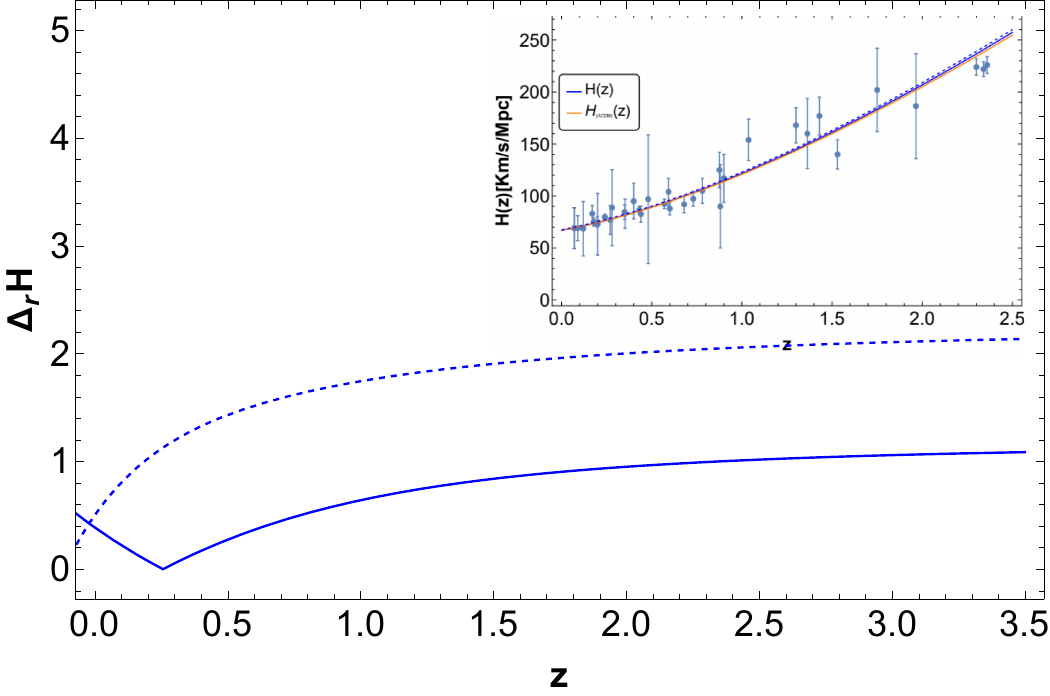}
    \caption{\scriptsize We present the evolution of the Hubble rate $H(z)$ and its relative difference $\Delta_r H(z) = 100\times |H-H_{\Lambda\text{CDM}}|/H_{\Lambda\text{CDM}}$ with respect to the $\Lambda\text{CDM}$ model, as functions of redshift, using the same initial conditions as shown in Figure \ref{fig:C.1.1}. This is complemented by the evolution of the Hubble rate $H_{\Lambda\text{CDM}}$ in the $\Lambda$CDM model and the Hubble data from references \cite{cao2018cosmological,farooq2013hubble}. We have employed the current value of the Hubble rate, $H_0=67.4$ km/(Mpc·s), from Planck 2018 \cite{Aghanim:2018eyx}.}
    \label{fig:C.1.5}
\end{figure}

Finally, for our third model, the numerical results are shown in FIGS. \ref{fig:C.1.7}, \ref{fig:C.1.1}, \ref{fig:C.1.3}, \ref{fig:C.1.4} and \ref{fig:C.1.5}.

In FIG. \ref{fig:C.1.7}, we show the evolution curves in the phase space for this ansatz. As for previous models, we have an attractor fixed point, which is a cosmological constant solution. This attractor solution with accelerated expansion has a variable position in the phase space, which depends on the model parameter related to the slope of the vector's potential. In FIG. \ref{fig:C.1.1}, we show the scaling behavior for early times.

On the other hand, in FIG. \ref{fig:C.1.3} we depict the evolution of the EoS of dark energy and the total EoS. As before, the effective dark energy fluid behaves as radiation field density at early times, whereas it behaves as a quintessence field at late times, in FIG. \ref{fig:C.1.4}, we observe that this model provides a deceleration-acceleration transition redshift closer to the value from the $\Lambda$CDM model, and therefore, it becomes compatible with observations \cite{Aghanim:2018eyx}. This result is also corroborated in FIG. \ref{fig:C.1.5}, where we plot the theoretical Hubble rate from our model along with the current Hubble rate data \cite{cao2018cosmological,farooq2013hubble}. 



\section{Thermodynamics}\label{Thermo}

The Universe, as a whole, can be seen as a classical thermodynamic system composed of interacting fluids. The temperature is defined through the Gibbs equation 
\be
T d\left(\frac{S}{N}\right)=d\left(\frac{\rho}{n}\right)+p\left(\frac{1}{n}\right),
\label{Gibbs_Eq}
\ee where $T$ is the temperature of the system and $S$ is total entropy per comoving volume and $N=n V=const$ with $V=a^3$, and $n$ is the number density.

\begin{figure}[!tbp]
    \centering
    \includegraphics[width=0.35\textwidth]{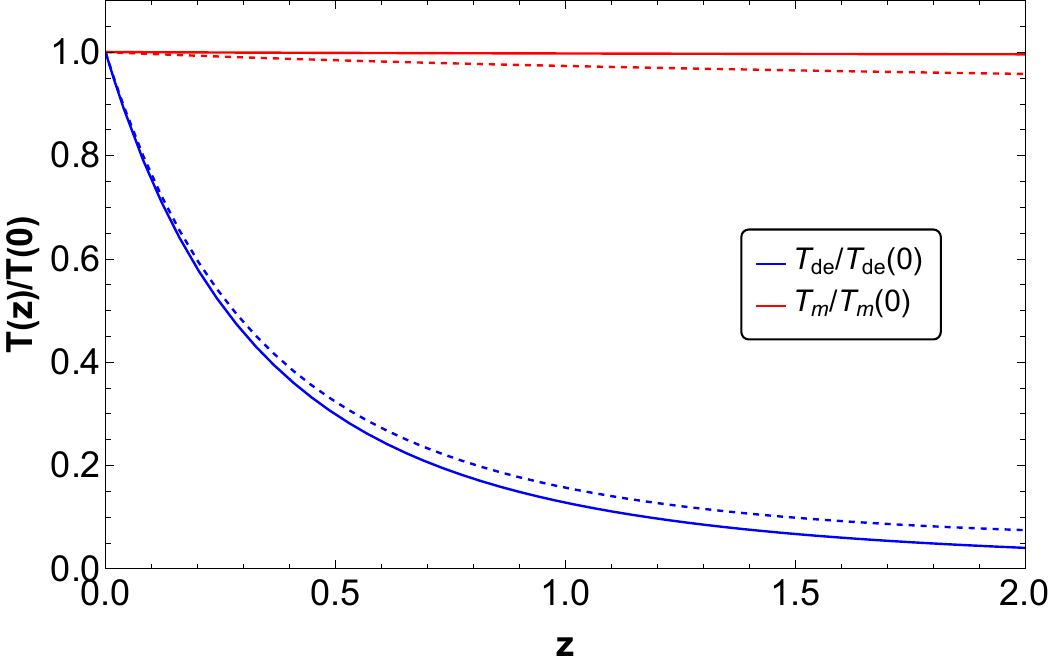}
    \caption{\scriptsize We depict the evolution of the temperature of dark energy and dark matter as a function of the redshift for interaction ``Case I'' using the same set of initial conditions used in Figure \ref{fig:A.1.1}.}
    \label{fig:T.1}
\end{figure}

\begin{figure}[!tbp]
    \centering
    \includegraphics[width=0.35\textwidth]{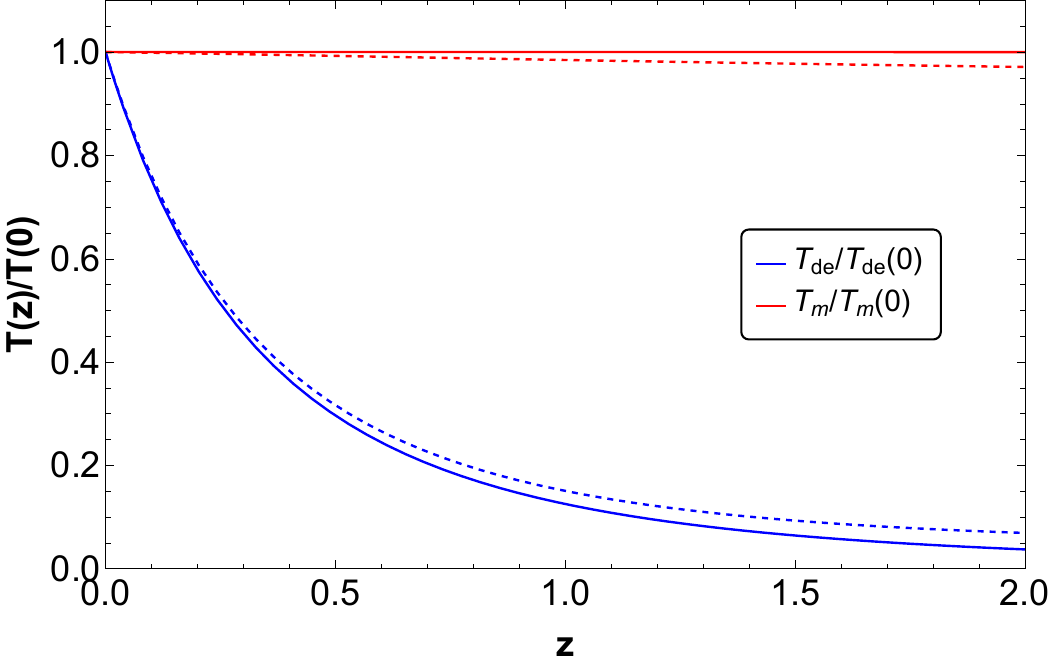}
    \caption{\scriptsize We depict the evolution of the temperature of dark energy and dark matter as a function of the redshift for interaction ``Case II'' using the same set of initial conditions used in Figure \ref{fig:B.1.1}.}
    \label{fig:T.2}
\end{figure}

\begin{figure}[!tbp]
    \centering
    \includegraphics[width=0.35\textwidth]{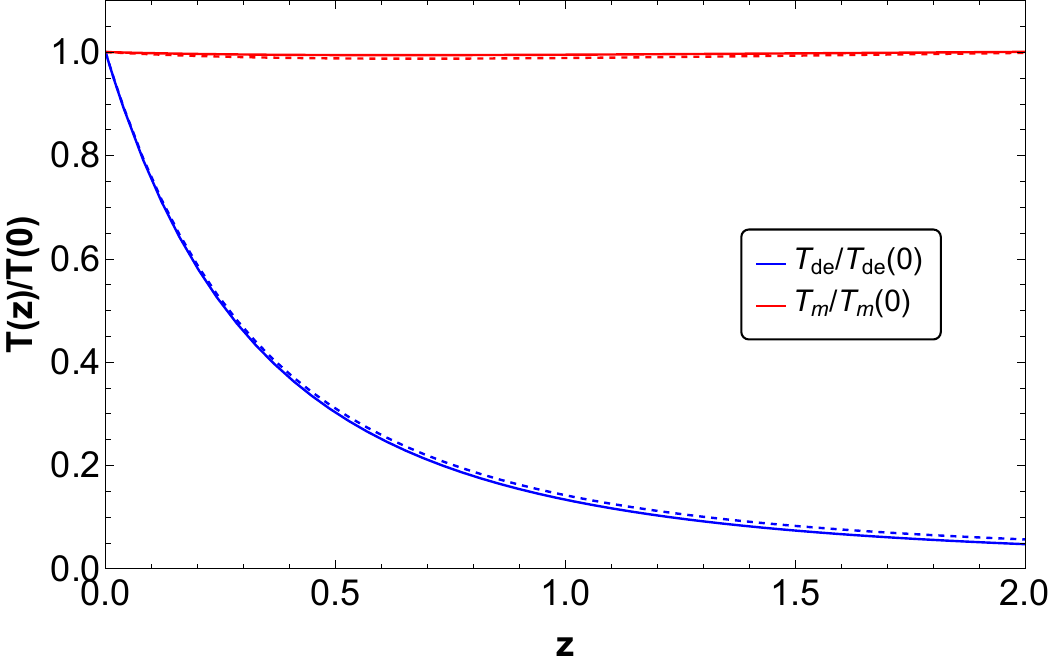}
    \caption{\scriptsize We depict the evolution of the temperature of dark energy and dark matter as a function of the redshift for interaction ``Case III'' using the same set of initial conditions used in Figure \ref{fig:C.1.1}.}
    \label{fig:T.3}
\end{figure}

\begin{figure}[!tbp]
    \centering
    \includegraphics[width=0.45\textwidth]{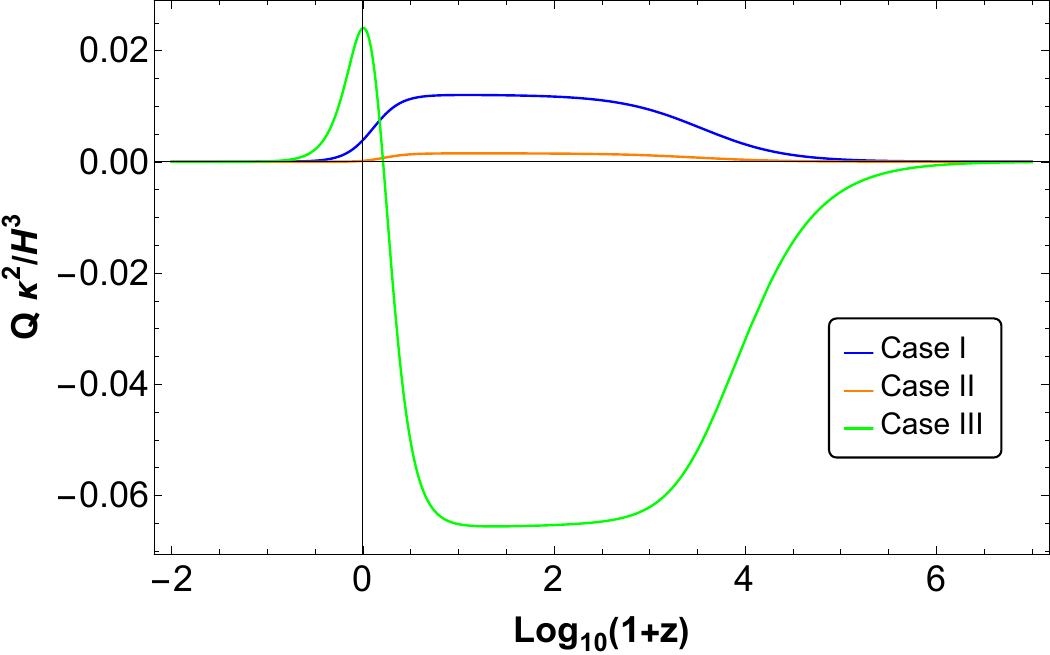}
    \caption{\scriptsize We depict the evolution of the interaction term $Q$ for each case considered. }
    \label{fig:T.4}
\end{figure}

This equation can be written as 
\be
T d\left(\frac{S}{N}\right)=d\left(\frac{d\rho}{n T}\right)-\left(\frac{\rho+p}{n^2 T}\right) dn. 
\ee Thus, the integrability condition
\be
\frac{\partial^2S}{\partial n\partial \rho}=\frac{\partial^2S}{\partial \rho \partial n}, 
\ee 
can allow us to obtain 
\be
n \frac{\partial T}{\partial n}+\left(\rho+p\right)\frac{\partial T}{\partial \rho}=T\frac{\partial p}{\partial \rho}.
\label{eq}
\ee
Since $T=T(n,\rho)$, and using the conservation laws
\bea
&& \dot{\rho}+3 H\left(\rho+p\right)=0, \\
&& \dot{n}+3 H n=0, 
\eea where $\rho$ and $p$ are the total energy and pressure densities, we obtain 
\bea
\dot{T}=-3 H\left[\left(\frac{\partial T}{\partial n}\right) n +\left(\frac{\partial T}{\partial \rho}\right)\left(\rho+p\right)\right]. 
\eea Using the relation \eqref{eq} we get
\be
\frac{\dot{T}}{T}=-3 H\left(\frac{\partial p}{\partial \rho}\right)_{n}.
\ee In the case of a barotropic perfect fluid with pressure $p=w(a) \rho$ we obtain 
\be
T(a)=T(a_{0}) exp\left[-3 \int_{a_{0}}^{a}{da \frac{w(a)}{a}}\right].
\label{Temp}
\ee For two interacting fluids (dark matter and dark energy) one can  write
\bea
&& \dot{\rho}_{de}+3H (1+w^{eff}_{de}) \rho_{de}=0, \\
&& \dot{\rho}_{m}+3H (1+w^{eff}_{m}) \rho_{m}=0,
\eea where it has been defined
\bea
&& w^{eff}_{de}=w_{de}+\frac{Q}{3 H\rho_{de}},\\
&& w^{eff}_{m}=w_{m}-\frac{Q}{3 H \rho_{m}},
\eea and then Eq. \eqref{Temp} gives us
{\small
\bea
\frac{T_{de}(z)}{T_{de}(0)} &=& \exp\left[3 \int_{0}^{z}{d \ln(1+z) w^{eff}_{de}(z)}\right],\\
\frac{T_{m}(z)}{T_{m}(0)} &=& (1+z)^{3w_{m}}\exp\left[- \int_{0}^{z}{d \ln(1+z)\left(\frac{Q}{H \rho_{m}}\right)}\right]. \nonumber\\
&&
\eea }
Thus, one can verify that for non-relativistic matter $w_{m}=0$, the temperature is constant in the absence of coupling. So, after specifying the coupling between dark matter and dark energy, we can evolve the temperature for both components as functions of redshifts. For the present interacting vector-like dark energy model, the time evolution of the effective EoS of dark energy $w^{eff}_{de}$ depends on both the magnitude of the vector field and the coupling to matter. Therefore, temperature behavior is determined by the dynamics of the vector field. In FIGS. \ref{fig:T.1}, \ref{fig:T.2},  and \ref{fig:T.3}, we depict the temperature behavior of matter and dark energy, parameterized in terms of the ratios $T_{de}(z)/T_{de}^{(0)}$ and $T_{m}(z)/T_{m}^{(0)}$, respectively, as functions of redshift $z$ for different initial conditions.  Here, we define $T_{m}^{(0)}\equiv T_{m}(z=0)$ and $T_{de}^{(0)}\equiv T_{de}(z=0)$.
Also, in FIG. \ref{fig:T.4}, we depict the behavior of the coupling function $Q(z)$ for the three models studied. 
From FIGS. \ref{fig:T.1}, \ref{fig:T.2}, and \ref{fig:T.3}, it is evident that the temperature of matter increases very slowly, while the temperature of dark energy rises more rapidly. Consequently, when we combine these findings with FIG.\ref{fig:T.4}, we observe an energy transference from dark energy to dark matter, indicating that dark energy possesses a negative heat capacity. Furthermore, it is verified that the second law of thermodynamics is satisfied during the regime $T_{m}<T_{de}$ for $Q>0$. From the Gibbs equation \eqref{Gibbs_Eq} and the continuity equations \eqref{rho_DE_Q} and \eqref{rho_DM_Q} one can demonstrate that $T_{de} dS_{de}=-Q Vdt$ and $T_{m} dS_{m}=Q Vdt$, where $S_{m}$ and $S_{de}$ represent the entropy of  matter and dark energy, respectively. Consequently, we can verify that $d(S_{de}+S_{m})=dS_{tot}=QV(1/T_{m}-1/T_{de})dt$, and the second law $d(S_{de}+S_{m})>0$, requires $T_{m}<T_{de}$ if $Q>0$, or $T_{m}>T_{de}$ if $Q<0$ \cite{Cardenas:2018nem}. In FIGS. \ref{fig:T.1}, \ref{fig:T.2}, and \ref{fig:T.3}, we observe that $T_{de}(z)/T_{de}^{(0)}<T_{m}(z)/T_{m}^{(0)}$. Thus, for $Q>0$, the validity of the second law requires $T_{m}<T_{de}$ \cite{Cardenas:2018nem}, and therefore $T_{m}^{(0)}<T_{de}^{(0)}$ (see Appendix \ref{appen_C}).

In the case of model III, we observe a sign change in $Q$, which suggests a change in the direction of the energy flux. This could imply a corresponding sign change in the heat capacity of dark energy, which must be in concordance with the temperature behavior in FIG. \ref{fig:T.3} (see Refs. \cite{Cardenas:2018nem,Lepe:2018owr}).

\section{Concluding Remarks}\label{Remarks}

In the present paper, we studied the cosmological dynamics of interacting vector-like dark energy. The vector-like dark energy refers to the ``cosmic triad", which consists of three identical vectors oriented in mutually orthogonal directions and thus preserving the space isotropy \cite{Armendariz-Picon:2004say}. Furthermore, we assumed an interaction between dark energy and dark matter. This interaction is described by the function $Q$, which is a function of the energy densities of both dark energy and dark matter, as well as the Hubble rate. Thus, we have studied several different models for $Q$ \cite{Wang:2016lxa}. 

We have shown that the cosmic triad can explain the dynamics of dark energy, providing a dark energy-dominated solution with accelerated expansion. These dark energy solutions are attractors fixed points, and thus, the system will reach them over a wide range of initial conditions.  Nevertheless, one can observe that the whole cosmological evolution is altered by the presence of interaction between the effective dark energy fluid and dark matter. In particular, we obtained new scaling solutions of radiation and dark matter, which originate from both the pure dynamics of the vector field and the matter-vector interaction function $Q$.  

It is important to note that although the critical point describing the dark energy-dominated era is an attractor, reachable through a wide set of initial conditions, fine-tuning is still required. This is to ensure consistency with the current estimations of cosmological parameters at redshift $z=0$, as well as to accurately reproduce the thermal history of the universe in line with observational data. The need for fine-tuning arises because the attractor critical point is only asymptotically achieved in the future, not precisely at $z=0$. To address this issue of finite-tuning, also known as the cosmological coincidence problem \cite{Copeland:2006wr}, the existence of an attractor scaling solution featuring accelerated expansion may be necessary. However, obtaining such solutions is inherently difficult.  Not all dark energy models offer this kind of solution \cite{Amendola:1999qq,Amendola:2006qi,Otalora:2013tba}, and in some cases, even when these solutions are present, the model may not successfully replicate the dark matter era \cite{Amendola:1999er}.

As the scaling solutions contribute to the existence of small amounts of dark energy in the radiation and matter era, they can give rise to notable physical outcomes \cite{Albuquerque:2018ymr,Ohashi:2009xw}. This specifically changes the Hubble rate during that era, leading to adjustments in the theoretical forecasts for the abundances of primordial light elements \cite{Ferreira:1997hj,Bean:2001wt}. Furthermore, the shape of the Cosmic Microwave Background (CMB) anisotropies spectrum is profoundly affected by such a scaling field \cite{Bean:2001wt}. The influence of this scaling field also extends to the development of the universe's large-scale structures, affecting the evolution of cosmic entities like galaxies and galaxy clusters \cite{Amendola:1999er}.

We also have studied the thermodynamics of our model, considering the interaction function $Q$. Using the Gibbs equation applied to an expanding universe, one can obtain general expressions for the temperature of dark matter and dark energy as functions of redshift in the presence of interaction between them \cite{Cardenas:2018nem}. We have analyzed three types of interaction functions $Q$ that depend on the energy density of dark energy and dark matter. Since the effective dark energy depends on the vector field and its dynamics, the interaction function $Q$ does well. In particular, we found for each interaction model the expressions for $Q$ in terms of the dynamical variables. By numerically solving the background equations and using the expressions for $Q$ in terms of the phase-space variables, we have depicted the behavior of the temperatures of dark energy and dark matter. Our results showed us that the temperature of matter increases very slowly, whereas the temperature of dark energy rises more quickly. Thus, we corroborated that there is a transference of energy from dark energy to dark matter, as dark energy has a negative heat capacity. In this way, we verified that the second law of thermodynamics is satisfied during the regime $T_{m}<T_{de}$ for $Q>0$ \cite{Cardenas:2018nem}, provided that $T^{(0)}_{m}<T^{(0)}_{de}$ However, it is important to note that measuring the temperature of dark matter and dark energy is extremely challenging \cite{Mirabolfathi:2013hda,Obreschkow:2012yb,Cardenas:2018nem}. Currently, the temperatures of these components remain unknown. Additionally, developing a suitable thermodynamic framework for the dark energy sector continues to be an area of active research due to the existing gaps in our understanding of its intrinsic nature. It is conceivable that future astronomical and cosmological observations will enhance our understanding of these elusive components , thereby enriching our knowledge of their thermodynamic evolution \cite{Wu:2022dgy,Zhao:2022bpd}. The analysis of the models studied through the interaction term (as shown in FIG. \ref{fig:T.4}) determines that the models include a change of sign for the dark energy interaction function. Since the interaction terms change their signs during evolution, our results indicate that today, dark energy is transferred to dark matter, but in the past, the transfer was the opposite \cite{arevalo2022dynamics, cid2021bayesian, rodriguez2020universe}. 


\begin{acknowledgments}
M. Gonzalez-Espinoza acknowledges the financial support of FONDECYT de Postdoctorado, N° 3230801. G. Otalora acknowledges Dirección de Investigación, Postgrado y Transferencia Tecnológica de la
Universidad de Tarapacá for financial support through
Proyecto UTA Mayor 4731-23. C. Rodriguez-Benites and M. Alva-Morales acknowledge the financial support of PE501082885-2023-PROCIENCIA. 

\end{acknowledgments}


\bibliography{bio}




\begin{appendix}
\section{Hubble's parameter data}\label{appen_B}

In this appendix, we present Hubble's parameter data for $0.01 < z < 2.360$:

\begin{table}[!htbp]
\caption{Hubble's parameter vs. redshift \& scale factor.}
\label{table:H(z)data}
\renewcommand{\tabcolsep}{0.7pc} 
\renewcommand{\arraystretch}{0.7} 
\begin{tabular}{@{}lllll}
\hline \hline
  $\;\; z$    &  $ H(z) \;$ ($\frac{km/s}{\text{Mpc}}$ ) &  Ref. \\
\hline
$0.07$      & $ \; \qquad 69     \pm 19.6 $      &   \cite{zhang2014} \\
$0.09$      & $ \; \qquad 69     \pm 12 $      & \cite{simon2005} \\
$0.100$     & $ \; \qquad 69     \pm 12 $      & \cite{simon2005} \\
$0.120$     & $ \; \qquad 68.6     \pm 26.2$       & \cite{zhang2014} \\
$0.170$     & $ \; \qquad 83     \pm 8$       & \cite{simon2005} \\
$0.179$     & $ \; \qquad 75     \pm 4$       & \cite{moresco2012} \\
$0.199$     & $ \; \qquad 75     \pm 5$        & \cite{moresco2012} \\
$0.200$     & $ \; \qquad 72.9     \pm 29.6$        &  \cite{zhang2014} \\
$0.270$     & $ \; \qquad 77     \pm 14$      & \cite{simon2005} \\
$0.280$     & $ \; \qquad 88.8     \pm 36.6$      & \cite{zhang2014} \\
$0.320$     & $ \; \qquad 79.2   \pm 5.6$     & \cite{cuesta2016}\\
$0.352$     & $ \; \qquad 83     \pm 14$      & \cite{moresco2012} \\
$0.3802$    & $ \; \qquad 83     \pm 13.5$      & \cite{moresco2012} \\
$0.400$     & $ \; \qquad 95     \pm 17$      & \cite{simon2005} \\
$0.4004$    & $ \; \qquad 77     \pm 10.2$      & \cite{moresco2012} \\
$0.4247$    & $ \; \qquad 87.1     \pm 11.2$      & \cite{moresco2012} \\
$0.440$     & $ \; \qquad 82.6   \pm 7.8$     & \cite{blake2012} \\
$0.4497$    & $ \; \qquad 92.8   \pm 12.9$     & \cite{moresco2012} \\
$0.470$     & $ \; \qquad 89   \pm 50$     & \cite{ratsim} \\
$0.4783$    & $ \; \qquad 80.9   \pm 9$     & \cite{moresco2012} \\
$0.480$     & $ \; \qquad 97     \pm 62$      & \cite{stern2010} \\
$0.570$     & $ \; \qquad 100.3  \pm 3.7$     & \cite{cuesta2016} \\
$0.593$     & $ \; \qquad  104   \pm 13$      & \cite{moresco2012} \\
$0.600$     & $ \; \qquad 87.9   \pm 6.1$     & \cite{blake2012} \\
$0.680$     & $ \; \qquad 92     \pm 8$       & \cite{moresco2012} \\
$0.730$     & $ \; \qquad 97.3   \pm 7 $      & \cite{blake2012} \\
$0.781$     & $ \; \qquad 105    \pm 12$      & \cite{moresco2012} \\
$0.875$     & $ \; \qquad 125    \pm 17$      & \cite{moresco2012} \\
$0.880$     & $ \; \qquad 90     \pm 40$      & \cite{stern2010} \\
$0.900$     & $ \; \qquad 117    \pm 23$      & \cite{simon2005} \\
$1.037$     & $ \; \qquad 154    \pm 20 $     & \cite{moresco2012} \\
$1.300$     & $ \; \qquad 168    \pm 17 $     & \cite{simon2005} \\
$1.363$     & $ \; \qquad 160    \pm 33.6$    & \cite{moresco2015}\\
$1.430$     & $ \; \qquad 177    \pm 18$      & \cite{simon2005}\\
$1.530$     & $ \; \qquad 140    \pm 14$      & \cite{simon2005}\\
$1.750$     & $ \; \qquad 202    \pm 40$      & \cite{simon2005}\\
$1.965$     & $ \; \qquad 186.5  \pm 50.4$    & \cite{moresco2015}\\
$2.340$     & $ \; \qquad 222    \pm 7 $      & \cite{delubac2014}\\
$2.360$     & $ \; \qquad 226    \pm  8$      & \cite{font-ribera2014}\\
\hline
\end{tabular}\\
 \end{table}

\vspace{20mm}
\section{Thermodynamics calculations\label{appen_C}}

From FIGS. \ref{fig:T.1}, \ref{fig:T.2}, and \ref{fig:T.3} we see that  
\be
\frac{T_{de}}{T^{(0)}_{de}}<\frac{T_{m}}{T^{(0)}_{m}},
\label{eq:B1}
\ee

where $T_{de}^{(0)}=T_{de}(z=0)$ and $T_{m}^{(0)}=T_{m}(z=0)$. Then, from \eqref{eq:B1} we obtain
\be
T_{de}<T_{m}\frac{T_{de}^{(0)}}{T_{m}^{(0)}}.
\ee

We know that for $Q>0$, the second law of thermodynamics requires $T_{m}<T_{de}$ \cite{Cardenas:2018nem}. Therefore, we have:
\be
T_{m}<T_{de}<T_{m}\frac{T^{(0)}_{de}}{T_{m}^{(0)}}.
\ee

Consequently, we get
\be
T_{m}<T_{m}\frac{T_{de}^{(0)}}{T_{m}^{(0)}}\ \Rightarrow\ T_{m}^{(0)}<T_{de}^{(0)}.
\ee
\\
\\
\\
\\
\\
\\
\\
\\
\\
\\
\\
\\
\\
\\
\\
\\
\\
\\

\end{appendix}

\end{document}